\documentclass[a4paper,twocolumn,11pt,accepted=2017-05-09]{quantumarticle}
\pdfoutput=1

\usepackage{graphicx,bm,amsmath}
\usepackage[usenames,dvipsnames]{xcolor}
\usepackage[colorlinks,bookmarks=false,citecolor=NavyBlue,linkcolor=Red,urlcolor=blue,
]{hyperref}

\usepackage[bbgreekl]{mathbbol}
\usepackage{xcolor}
\usepackage[normalem]{ulem}
\usepackage{algorithm}
\usepackage{braket}
\usepackage{algpseudocode}
\usepackage{comment}
\usepackage{amsthm}
\usepackage{enumerate}

\def\doi{http://dx.doi.org/}

\newcommand{\be}{\begin{equation}}
\newcommand{\ee}{\end{equation}}
\newcommand{\bec}{\begin{equation*}}
\newcommand{\eec}{\end{equation*}}
\newcommand{\bea}{\begin{eqnarray}}
\newcommand{\eea}{\end{eqnarray}}

\newcommand{\titleinfo}{Magic in generalized Rokhsar-Kivelson wavefunctions}

\newcommand{\Tr}{\text{Tr}}   

\begin{document}

\title{\titleinfo}
\author{Poetri Sonya Tarabunga}
\affiliation{The Abdus Salam International Centre for Theoretical Physics (ICTP), Strada Costiera 11, 34151 Trieste,
Italy}
\affiliation{International School for Advanced Studies (SISSA), via Bonomea 265, 34136 Trieste, Italy}
\affiliation{INFN, Sezione di Trieste, Via Valerio 2, 34127 Trieste, Italy}
\author{Claudio Castelnovo}
\affiliation{TCM Group, Cavendish Laboratory, University of Cambridge, Cambridge CB3 0HE, UK}

\begin{abstract}
Magic is a property of a quantum state that characterizes its deviation from a stabilizer state, serving as a useful resource for achieving universal quantum computation e.g., within schemes that use Clifford operations. 
To date little is known about properties and behaviour of magic in many body quantum systems. 
In this work, we study magic, as quantified by the stabilizer Renyi entropy, in a class of models known as generalized Rokhsar-Kivelson systems, i.e., Hamiltonians that allow a stochastic matrix form (SMF) decomposition. The ground state wavefunctions of these systems can be written explicitly throughout their phase diagram, and their properties can be related to associated classical statistical mechanics problems, thereby allowing powerful analytical and numerical approaches that are not usually available in conventional quantum many body settings. 
As a result, we are able to express the SRE of integer Renyi index $n>1$ in terms of wave function coefficients that can be understood as a free energy difference of related classical problems. 
We apply this insight to a range of quantum many body SMF Hamiltonians, which affords us to study numerically the SRE of large high-dimensional systems, unattainable with existing tensor network-based techniques, and in some cases to obtain analytical results. 
We observe that the behaviour of the SRE is relatively featureless across quantum phase transitions in these systems, although it is indeed singular (in its first or higher order derivative, depending on the first or higher order nature of the transition). On the contrary, we find that the maximum of the SRE generically occurs at a cusp away from the quantum critical point, where the derivative suddenly changes sign. 
Furthermore, 
we compare the SRE and the logarithm of overlaps with specific stabilizer states, asymptotically realised in the ground state phase diagrams of these systems. We find that they display strikingly similar behaviors, which in turn establish rigorous bounds on the min-relative entropy of magic.
\end{abstract}

\maketitle
%
%

\section{Introduction}

Stabilizer states are an important class of quantum states in quantum information theory~\cite{gottesman1997stabilizer,nielsen2002quantum}. They have very rich structures~\cite{vedral1997quantifying,horodecki2009quantum} and can be highly entangled~\cite{smith2006typical,gutschow2010entanglement,preskill2012quantum,harrow2017quantum}. However, it is well-known by the Gottesman-Knill theorem (and its subsequent extensions) that quantum computation using only stabilizer states and Clifford circuits can be efficiently simulated on a classical computer~\cite{gottesman1997stabilizer,gottesman1998heisenberg,gottesman1998theory,aaronson2004improved}. In quantum computation using the state injection scheme~\cite{bravyi2005UniversalQuantumComputation,campbell2017roads,bravyi2012magic}, universal quantum computation is achieved by injection of states outside the stabilizer set, known as magic states, while keeping the set of operations restricted to Clifford operations. Magic thus serves as a fundamental resource that would be required to outperform classical simulations. To quantify the amount of magic resource in a quantum state, the notion of a magic measure has been introduced within the framework of resource theory~\cite{chitambar2019}. These measures assess the amount of resource a state can provide in quantum computation by state injection scheme, offering insights into the computational power and quantum capabilities of different states. Most measures that have been introduced require optimization procedures to compute them (see, e.g., Refs.~\cite{bravy2016improved,bravy2016trading,Howard2017,Heinrich2019,Seddon2021}), and are thus difficult to evaluate beyond a few qubits~\cite{hamaguchi2023handbook}. More recently, a computable measure of magic has been introduced, called Stabilizer Renyi Entropy (SRE)~\cite{leone2022stabilizer}, which is relatively easier to compute as it is expressed only in terms of expectation values of Pauli strings. Nonetheless, examples of analytical results for the SRE remain few and far between, and computational cost often limits numerical studies to relatively small systems. 

In recent years, there is an increasing interest in characterizing the role of magic in the ground states of quantum many-body systems~\cite{liu2022,white2021,Sarkar2020,oliviero2022ising, odavić2022complexity,haug2023quantifying,haug2023stabilizer,lami2023quantum,tarabunga2023manybody,tarabunga2023critical,tarabunga2024nonstabilizerness,frau2024nonstabilizerness,chen2023magic} -- namely, `how far' such states are from being stabilizer states. Notable progress has also been made in its experimental measurements~\cite{oliviero2022measuring, haug2023scalable, haug2023efficient, tarabunga2023manybody,tirrito2023,turkeshi2023measuring}. The study of quantum information concepts in the realm of many-body theory has a long history, with a prominent example being entanglement, which has proven to be a powerful tool for investigating various many-body phenomena~\cite{amico2008,eisert2010}. The comprehensive exploration of many-body magic thus represents a promising avenue that holds significant relevance for characterizing the quantum complexity of different phases of matter. This includes understanding the computational cost of simulating specific phases and assessing their computational capabilities. In a more practical context, this would also provide insights into the potential of many-body states to be an input in magic state distillation~\cite{bao2022}, for applications in quantum information processing. 

The investigation of many-body magic has been enabled by the recent development of numerical methods based on tensor networks to efficiently compute the SRE~\cite{haug2023quantifying,haug2023stabilizer,lami2023quantum,tarabunga2023manybody,tarabunga2024nonstabilizerness}. In particular, these studies have suggested a connection between magic and criticality in one-dimensional quantum systems. Notwithstanding these interesting developments, the cost of computing the SRE remains very high, often restricting the study to simple one-dimensional systems. Further studies of highly entangled states, such as higher-dimensional systems, appear to be practically out of reach with current methods (see however Ref.~\cite{tarabunga2023manybody}).

In this work, we introduce an approach to compute the SRE with integer Renyi index $n>1$ in many-body wavefunction, by expressing it in terms of wavefunction coefficients that make it amenable to computation using Monte Carlo sampling (provided the wavefunction can be gauged to have non-negative coefficients). 

We apply this approach to a class of models known as generalized Rokhsar-Kivelson systems~\cite{Henley2004,Ardonne2004}, or Hamiltonians that allow a stochastic matrix form (SMF) decomposition~\cite{Castelnovo2005}. The ground state wavefunctions of these systems can be written explicitly throughout their phase diagram, and their properties can be related to associated classical statistical mechanics problems in thermodynamic equilibrium at temperature $T$, which plays the role of a parameter in the phase diagram of the original quantum systems. This correspondence allows powerful analytical and numerical approaches to be deployed, that are not usually available in conventional quantum many body settings~\footnote{We note that a study of the SRE in related Rokhsar-Kivelson-sign wavefunctions was presented in Ref.~\cite{piemontese2023}.}. 

Since the early work of Rokhsar and Kivelson~\cite{Rokhsar_Kivelson}, SMF Hamiltonian and wavefunctions have appeared in many different physics contexts. Of late, a resurgence of attention has derived from the fact that they can be naturally realised using tensor networks and PEPS~\cite{Verstraete2006}, and they can be implemented in measurement-prepared quantum states and (monitored) quantum circuits~\cite{Schwarz2012,zhu2019,lee2022decoding,zhu2022nishimoris,chen2023realizing,zhu2023}. In this context, the magic of SMF wavefunctions thus directly quantifies the amount of non-Clifford resources required to prepare these systems in the circuits.

We are able to express the SRE of SMF wavefunctions in terms of a free energy difference of related classical problems, which can then be efficiently computed by thermodynamic integration. We apply this insight to a range of quantum many body SMF Hamiltonians, which affords us to study numerically the SRE of large high-dimensional systems, unattainable using existing tensor network-based techniques, and in some cases we obtain explicit analytical results. 

We observe that the behaviour of the SRE is relatively featureless across quantum phase transitions in these systems~\cite{Castelnovo2010}, although it is indeed singular in its first or higher order derivative, depending on the first or higher order nature of the transition. On the contrary, we find that the maximum of the SRE generically occurs at a cusp away from the quantum critical point, where the derivative suddenly changes sign. Furthermore, we compare the SRE to the logarithm of overlaps with specific stabilizer states, that are asymptotically realised in the ground state phase diagrams of these systems. We find that they display strikingly similar behaviors, which in turn establish rigorous bounds on the min-relative entropy of magic.

The rest of the paper is structured as follows. In Sec.~\ref{sec:smf} and Sec.~\ref{sec:sre} we give a brief review of SMF Hamiltonians and SRE, respectively. We then state our general results in Sec.~\ref{sec:general_results} about the SRE and its upper bounds in SMF systems. In Sec.~\ref{sec:models} we then present a study of a range of models, encompassing the Ising ferromagnet in 1D, 2D, 3D, and infinite dimensions; the $J_1-J_2$ model on the square lattice; the triangular Ising antiferromagnet; and the Edwards-Anderson model on the cubic lattice. Finally, we conclude in Sec.~\ref{sec:conclusions}. 
%
%

\section{Brief review of Stochastic Matrix Form (SMF) Hamiltonians}
\label{sec:smf}
The stochastic matrix form wavefunctions, dependent on the parameter $\beta$, are given by~\cite{Henley2004,Ardonne2004, Castelnovo2005} 
\begin{equation} \label{eq:smf_wf}
    | \psi_{\rm SMF} \rangle = \frac{1}{\sqrt{Z(\beta)}} \sum_{\sigma} e^{-\beta E_\sigma/2} | \sigma \rangle
    \, ,
\end{equation}
where
\begin{equation}
    Z(\beta) = \sum_\sigma e^{-\beta E_\sigma}
    \, .
\end{equation}
$Z(\beta)$ can be seen as a classical partition function at temperature $T=1/\beta$. One can design a quantum Hamiltonian for arbitrary choice of $E_\sigma$, such that $| \psi_{\rm SMF} \rangle$ is the ground state of the Hamiltonian. In particular, for a locally interacting $E_\sigma$, the Hamiltonian also contains only local interactions. The quantum Hamiltonian is said to be SMF decomposable \cite{Castelnovo2005}. The equal-time correlation function of diagonal operators of SMF wavefunctions are given by the equal-time correlations functions in the associated classical systems in thermal equilibrium. It follows that the ground state phase diagram of the quantum Hamiltonian contains the thermal phase diagram of the classical system in thermal equilibrium. Since the wave function coefficients are known exactly by design, the wave function can be sampled with classical Monte Carlo simulations of the corresponding classical system.
%
%

\section{Stabilizer Rényi entropy} 
\label{sec:sre}
Stabilizer Rényi Entropies (SREs) are a measure of nonstabilizerness recently introduced in Ref.~\cite{leone2022stabilizer}. For a pure quantum state $|\psi \rangle$ of a system of $N$ qubits (equivalently, spin-$1/2$ degrees of freedom), SREs are expressed in terms of the expectation values of all Pauli strings in $\mathcal{P}_N$: 
\be \label{eq:SRE_def}
M_n \left( |\psi \rangle \right)= \frac{1}{1-n} \log \left \lbrace \sum_{P \in \mathcal{P}_N} \frac{\langle \psi | P | \psi \rangle^{2n}}{2^N} \right \rbrace \, .
\ee
Eq.~\eqref{eq:SRE_def} can be seen as the Rényi-$n$ entropy of the classical probability distribution $\Xi_P (|\psi\rangle)=\langle \psi | P | \psi \rangle^2 / 2^N$, also known as the characteristic function~\cite{Gross2021}.
The SREs have the following properties~\cite{leone2022stabilizer}: (i) faithfulness, i.e., $M_n(|\psi \rangle)=0$ iff $|\psi \rangle \in \text{STAB}$; (ii) stability under Clifford unitaries $C \in \mathcal{C}_N$, i.e., $M_n(C|\psi \rangle )=M_n(|\psi \rangle)$; and (iii) additivity, i.e., $M_n(|\psi \rangle_{A} \otimes|\psi \rangle_{B})=M_n(|\psi \rangle_{A})+M_n(|\psi \rangle_{B})$. The SREs are thus a good measure of magic from the point of view of resource theory.

The SRE is related to another magic monotone called the min-relative entropy of magic, defined as
\begin{equation}
    D_{min}(|\psi \rangle) = -\log F_{\text{STAB}}(|\psi \rangle)
\end{equation}
where $F_{\rm STAB}$ is the stabilizer fidelity defined as
\begin{equation}
    F_{\rm STAB}(|\psi \rangle) = \max_{|\phi \rangle \in \text{STAB}} | \langle \phi |\psi \rangle|^2
    \, .
\end{equation} 
The following inequality holds~\cite{haug2023stabilizer}
\begin{equation} \label{eq:ineq_dmin}
    M_n(|\psi \rangle) \leq   \frac{2n}{n-1} D_{\text{min}}(|\psi \rangle) \, ,  \quad n \geq 1
    \, .
\end{equation}

In particular, setting $n=2$, and defining $D(|\psi \rangle,|\phi \rangle) = -\log | \langle \phi |\psi \rangle|^2$, one finds that 
\begin{equation} \label{eq:ineq_dmin2}
    M_2(|\psi \rangle) \leq 4 D_{\text{min}}(|\psi \rangle) \leq 4D(|\psi \rangle,|\phi \rangle)
    \, , 
\end{equation}
for any $|\phi \rangle \in \text{STAB}$.

Because the SRE is defined in terms of the characteristic function $\Xi_P(|\psi\rangle)$, it is natural to estimate it through statistical sampling of $\Xi_P(|\psi\rangle)$. Indeed, this was the core of previous numerical methods that have been introduced to compute the SRE~\cite{haug2023stabilizer,lami2023quantum,tarabunga2023manybody}, which are so far limited to tensor network techniques. To obtain reliable statistics, a large number of samples is required, often resulting in computations being restricted to small bond dimension. This limitation poses a challenge when studying highly entangled systems, such as higher dimensional ones. Unfortunately, the existing methods are not directly suitable for Quantum Monte Carlo approaches due to the inherent difficulty in evaluating the expectation values of high-weight Pauli strings within.
%
%

\section{Magic in SMF ground states
\label{sec:general_results}}
In this section we show how the special form of the ground state wave functions of SMF systems allows for analytical and numerical routes into the calculation of their magic, that are not afforded to general quantum many body systems. In doing so, we develop the machinery that will later be used to study a broad range of model systems, to gain insight in the behaviour of this intriguing quantity. 
%
%

\subsection{Stabilizer Rényi entropy}
\label{sec:SMFsre}
Consider a $N-$qubit wave-function $| \psi \rangle = \sum_{{\sigma}} c_{\sigma} | \sigma \rangle $, where $c_\sigma  = \langle \sigma | \psi \rangle$ and $\sigma$ labels all tensor product basis states (e.g., the $\sigma_i^z$ basis for a spin-$1/2$ system, $i=1,...,N$). 
We firstly show, as detailed in Appendix~\ref{sec:proof}, that the SRE $M_2$ can be expressed as follows 
\begin{equation} \label{eq:sre}
    \begin{split}
    \exp(-M_2) = 
    \!\!\!\!\sum_{\sigma^{(1)},\sigma^{(2)},\sigma^{(3)},\sigma^{(4)}}\! 
    \left[ \vphantom{\sum}
    c_{\sigma^{(1)}} c_{\sigma^{(2)}} c_{\sigma^{(3)}}  c_{\sigma^{(1)}\sigma^{(2)}\sigma^{(3)}} 
    \right.
    \\
    \left. \vphantom{\sum}
    c^{*}_{\sigma^{(1)} \sigma^{(2)} \sigma^{(4)}} c^{*}_{\sigma^{(1)} \sigma^{(3)} \sigma^{(4)}} c^{*}_{\sigma^{(2)}\sigma^{(3)}\sigma^{(4)}} 
    c^{*}_{\sigma^{(4)}} 
    \right] \, , 
    \end{split}
\end{equation}
where $c_{\sigma^{(1)}\sigma^{(2)}\sigma^{(3)}}$ denotes the coefficient $c_{\tilde\sigma}$ corresponding to the tensor product label $\tilde\sigma$ given by the point product of $\sigma^{(1)}$, $\sigma^{(2)}$, and $\sigma^{(3)}$: $\tilde\sigma_i = \sigma^{(1)}_i\sigma^{(2)}_i\sigma^{(3)}_i$, $\forall\,i=1,\ldots,N$. Similarly for the other equivalent terms in Eq.~\eqref{eq:sre}. We can imagine $\sigma^{(a)}, \: a=1,2,3,4$ as four copies of the $N$ qubits. Similar expressions can be obtained for SREs of integer index $n>2$. 

For the SMF systems introduced in Sec.~\ref{sec:smf}, and in particular their ground state wavefunctions, these known expressions for the magic can be further manipulated upon substituting $c_\sigma = e^{-\beta E_\sigma/2}/\sqrt{Z}$ in Eq.~\eqref{eq:sre} to obtain: 
\begin{equation} \label{eq:sre_smf}
    M_2 = -\log{\frac{Z_M}{Z^4}} 
\end{equation} 
where
\begin{equation} \label{eq:z_m}
    \begin{split}
    Z_M &= \sum e^{-\beta E_M} \\
    &= {\sum_{\scriptsize
    \begin{array}{l}
    \sigma^{(1)},\sigma^{(2)}, 
    \\ 
    \sigma^{(3)},\sigma^{(4)}
    \end{array}
    }} 
    \exp\left[-\beta \sum_a E_{\sigma^{(a)}} 
    -\beta \sum_a E_{\prod_{b \neq a} \sigma^{(b)}}\right]
    \, . 
    \end{split}
\end{equation}
One can interpret $Z_M$ as a classical partition function constructed from four copies of the original classical degrees of freedom. The second term in the square bracket describes the energy of a configuration obtained from the spin product of three out of four copies. The expression in Eq.~\eqref{eq:sre_smf} can thus be seen as (proportional to) the difference between the free energy of the classical system described by $Z_M$ and four non-interacting copies of the original classical system described by $Z$. 

This manipulation is not only interesting from a conceptual point of view, but also from a pragmatic one: it provides a new angle to compute the magic of a quantum (SMF) state using classical statistical mechanics tools in the same number of dimensions. As we demonstrate later, it affords us the ability to access significantly larger system sizes and higher dimensional lattices than previously possible~\cite{haug2023stabilizer,lami2023quantum,tarabunga2023manybody}. 

In practice, rather than computing ratios of partition functions or differences in free energies, it is convenient to notice that the derivative of $M_2$ with respect to temperature reduces to: 
\begin{equation} \label{eq:m2_derivative}
    \frac{dM_2}{dT} = \frac{4 \langle E \rangle - \langle E_M \rangle_M}{T^2}
    \, . 
\end{equation}
Thus, $M_2$ can be more efficiently obtained by computing the energies $\langle E \rangle$ and $\langle E_M \rangle_M$ and then proceding to integrate the r.h.s. of Eq.~\eqref{eq:m2_derivative}.
%
%

\subsection{Upper bound of $M_2$} 
\label{sec:bound_m2}
As discussed in Sec.~\ref{sec:sre}, $M_2$ is bounded from above by $D(|\psi \rangle,|\phi \rangle) = -\log | \langle \phi |\psi \rangle|^2$ for any $|\phi \rangle \in \text{STAB}$. 
In the following sections, we compute $D(|\psi \rangle,|\phi \rangle)$ for several ad hoc choices of states $|\phi \rangle$, specific to the system being considered. Once again, in the case of SMF wavefunctions, these overlaps can be reduced to classical statistical mechanical objects, amenable to corresponding analytical or numerical estimates. 

Here, we illustrate the procedure to compute $D(|\psi \rangle,|\phi \rangle)$ in a couple of cases that will often be used in the following. 
Consider for example $|\phi \rangle =  |+++... \rangle$, where $| + \rangle = \frac{| \uparrow \rangle + | \downarrow \rangle}{2}$ (i.e., a spin-$1/2$ state polarized in the $x$ direction). In the context of SMF wavefunctions, this is the ground state at $\beta=0$. We denote $D_x(|\psi \rangle) = D(|\psi \rangle,|\phi \rangle = |+++... \rangle)$. From the overlap
\begin{align}
    \langle \psi(\beta) | \psi(\beta=0) \rangle &= \frac{1}{\sqrt{Z(\beta)2^N} } \sum_{\sigma, \sigma'} e^{-\beta E_\sigma/2} \langle \sigma' | \sigma \rangle \\
    &= \frac{1}{\sqrt{Z(\beta)2^N} } \sum_{\sigma} e^{-\beta E_\sigma/2}  \\
    &= \frac{Z(\beta/2)}{\sqrt{Z(\beta)2^N} }
    \, , 
\end{align}
we then obtain
\begin{equation} \label{eq:Dx}
    D_x(|\psi(\beta) \rangle) = -\log \frac{Z(\beta/2)^2}{Z(\beta)2^N} 
    \, . 
\end{equation}

Another useful example is $\vert\phi \rangle = (\vert \uparrow \uparrow \uparrow...\rangle + \vert\downarrow \downarrow \downarrow...\rangle)/\sqrt{2}$. In the ferromagnetic Ising model, this is the ground state at $T=0$ (namely, the exact ground state, ignoring spontaneous symmetry breaking effects). We denote $D_{zz}(|\psi \rangle) = D(|\psi \rangle,|\phi \rangle = (|\uparrow \uparrow \uparrow...\rangle + |\downarrow \downarrow \downarrow...\rangle)/\sqrt{2})$. From the overlap
\begin{align}
    \langle \psi(\beta) | \phi \rangle = \sqrt{\frac{2}{Z(\beta)}}  e^{-\beta E_{zz}/2}
    \, ,
\end{align}
where $E_{zz}$ is the energy of the configuration $\sigma_i = 1(-1)$, for all $i$, we obtain
\begin{equation} \label{eq:Dzz}
    D_{zz}(|\psi(\beta) \rangle) = \beta E_{zz} -\log \frac{2 }{Z(\beta)} 
    \, . 
\end{equation}

In all cases, the overlaps $D(|\psi \rangle,|\phi \rangle)$ are related to the partition function of the classical model. Therefore, similarly to $M_2$, we compute them using direct thermodynamics integration.
%
%

\section{SMF models
\label{sec:models}}
Armed with the tools developed above, we proceed to investigate a broad range of models, in the attempt to deepen our understanding of stabilizerness and magic in many body systems -- albeit of the fine tuned SMF kind -- and its relation to quantum phase transitions. 
For this purpose, we consider in the first instance quantum Ising ferromagnets in 1D, 2D, 3D, and infinite dimensions (mean field); we also consider the $J_1$-$J_2$ model tuned to exhibit a first order phase transition, for comparison. 
We then move on to more exotic examples, such as the quantum triangular Ising antiferromagnet (which is fully frustrated in the SMF realisation) and the Edwards-Anderson model (exemplifying a spin glass transition in the droplet picture). 

Since the SREs are generally an extensive quantity, we focus on the SRE densities $M_n/N$, where $N=L^d$ is the total number of sites in the system with linear size $L$ and dimensionality $d$. For the numerical simulations at finite volume, we impose periodic boundary conditions.

%
%

\subsection{1D SMF Ising ferromagnet}
The quantum SMF Hamiltonian for the 1D Ising ferromagnet is related to the classical 1D Ising model with energy 
\begin{equation}
    E_\sigma = - \sum_{i} \sigma_i \sigma_{i+1}
    \, , 
\end{equation}
where the spins $\sigma_i$ live on a chain, and it reflects its thermodynamic behaviour. In particular, there is no phase transition and the system orders only in the limit $\beta \to \infty$ (we stress that $\beta$ plays the role of inverse temperature for the classical system whereas it is merely a tunable parameter in the SMF quantum Hamiltonian, which is considered to be at zero temperature in its ground state, for the purpose of this work).

Despite its simplicity, this model serves as a useful warmup example and the magic can be computed analytically using Eq.~\eqref{eq:sre_smf}.
\begin{figure} 
    \centering
    \includegraphics[width=1\linewidth]{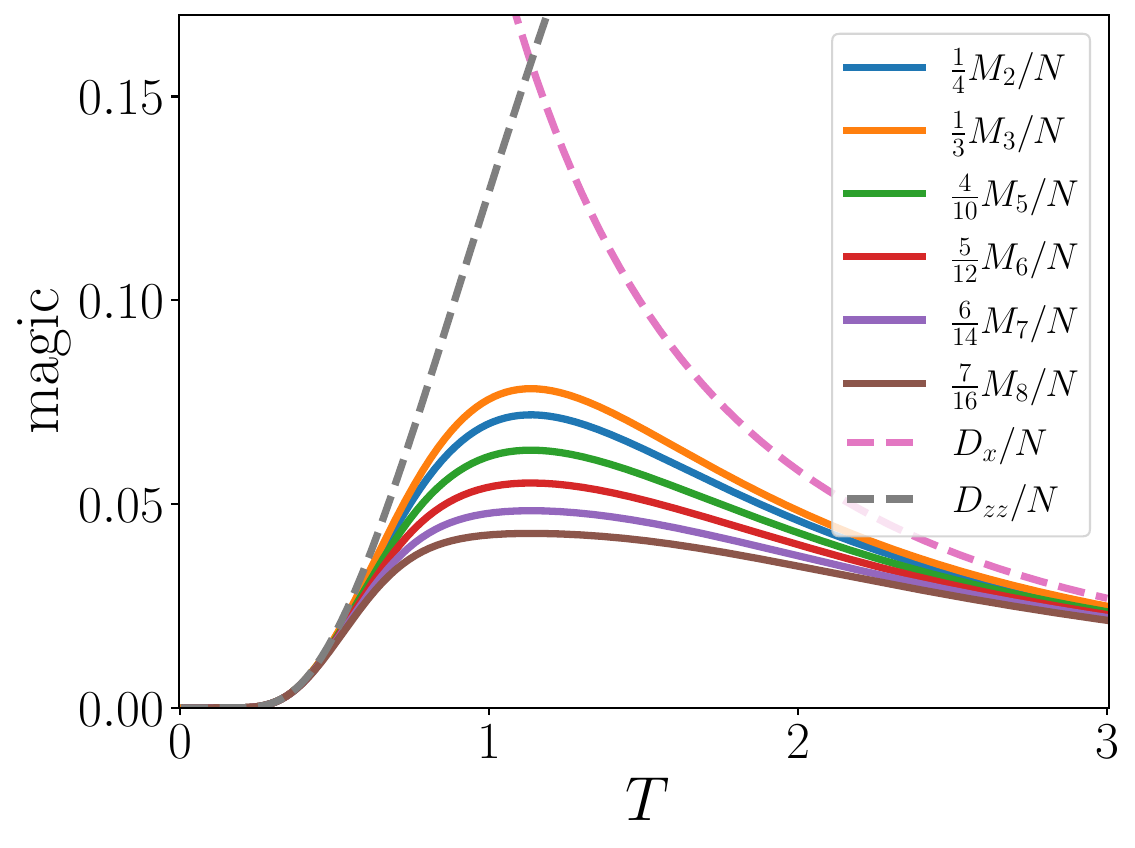}
    \caption{Behaviour of various measures of magic, including two stabilizer bounds, for the SMF 1D Ising ferromagnet. For each $n$, we plot $\frac{n-1}{2n} M_n / N$, which are upper bounded by $D_x$ and $D_{zz}$ by Eq. \eqref{eq:ineq_dmin}. }
    \label{fig:1dising}
\end{figure}
Indeed, we recall that the partition function of the 1D (nearest-neighbour) Ising model can be computed with transfer matrix techniques: 
\begin{equation}
\begin{split}
    Z &= \sum_{\sigma} e^{\beta \sum_{i} \sigma_i \sigma_{i+1}} \\
     &= \sum_{\sigma} e^{\beta \sigma_1 \sigma_{2}} e^{\beta \sigma_2 \sigma_{3}} ... e^{\beta \sigma_L \sigma_{1}} \\
     &= \sum_{\sigma} V_{\sigma_1,\sigma_2} V_{\sigma_2,\sigma_3} ... V_{\sigma_L,\sigma_1} \\
     &= \textrm{Tr}(V^L)
     \, ,
\end{split}
\end{equation}
where $V_{\eta,\eta'}=e^{\beta \eta \eta'}$ is a $2 \times 2$ matrix. The eigenvalues of $V$ are $\lambda_1=2 \cosh{\beta}$ and $\lambda_2=2 \sinh{\beta }$. Thus,
\begin{equation} \label{eq:z_1d}
    \begin{split}
        Z=\textrm{Tr}(V^L)=\lambda_1^L + \lambda_2^L =  [2\cosh{\beta }]^L + [2\sinh{\beta }]^L
        \, .
    \end{split}
\end{equation}

We can similarly compute $Z_M$ by interpreting the four layers of the 1D chain as a 1D chain with 16 states per site: $Z_M= \Tr (V_M^L)$, where $V_M$ is a $16 \times 16$ matrix.

More generally, for an integer index $n > 1$, the transfer matrix $V_{M,n}$ is a $2^{2n} \times 2^{2n}$ matrix with elements
\begin{eqnarray}
&&\!\!\!\!\!\!\!\!\!\!
V_M\left(\{\eta^{(1)},...,\eta^{(2n)}\},\{\eta'^{(1)},...,\eta'^{(2n)} \}\right) =
\nonumber \\ 
&&
\exp\left[
\frac{\beta}{2} \sum_a {\eta^{(a)}}{\eta'^{(a)}} 
+
\frac{\beta}{2} \sum_a \prod_{b \neq a} \eta^{(b)} \eta'^{(b)}
\right]
\, .
\end{eqnarray}
To compute $Z_{M,n} = \Tr(V_{M,n}^L)$ we work in the thermodynamic limit $L\to\infty$, where we only need to find the largest eigenvalue of $V_{M,n}$. One can verify that the column vector with all elements equal to $1$ is an eigenvector of $V_{M,n}$. By the Perron-Frobenius theorem, the corresponding eigenvalue is the unique largest eigenvalue: 
\begin{equation} \label{eq:largest_eigenvalue}
    \lambda_n = \frac{2^{2n}+[2\cosh(\beta)]^{2n} + [2\sinh(\beta)]^{2n}}{2}
    \, .
\end{equation}
Finally, using Eq.~\eqref{eq:largest_eigenvalue} and~\eqref{eq:z_1d}, we find
\begin{equation}
    M_n/N = \frac{1}{1-n} \log \frac{1+ \cosh(\beta)^{2n} + \sinh(\beta)^{2n}}{2 \cosh(\beta)^{2n}}
    \, , 
\end{equation}
for integer $n>1$.

Furthermore, $D_x$ and $D_{zz}$ can be computed directly by plugging in the partition function Eq.~\eqref{eq:z_1d} into Eq.~\eqref{eq:Dx} and~\eqref{eq:Dzz}, respectively. In Fig.~\ref{fig:1dising}, we show the SREs, $D_x$, and $D_{zz}$ of the SMF 1D Ising model, observing the expected asymptotic agreement in the limits $T\to0$ and $T\to\infty$. 

To avoid confusion, we remark here that the SMF Hamiltonian and corresponding GS wavefunction related to the classical 1D Ising ferromagnet are strikingly different from the conventional 1D Ising ferromagnet in a transverse field. Most notably, in the SMF case there is no phase transition and ordering occurs only asymptotically in the limit $T \to 0$. 
%
%

\subsection{2D SMF Ising ferromagnet
\label{sec:Ising2D}}
The quantum SMF Hamiltonian of the 2D Ising ferromagnet is related to the 2D classical Ising model with energy 
\begin{equation} \label{eq:2dising}
    E_\sigma = - \sum_{\langle ij \rangle} \sigma_i \sigma_j
    \, ,
\end{equation}
where the spins $\sigma_i$ are taken without loss of generality to live on the 2D square lattice. There is a phase transition between the ferromagnetic and the paramagnetic phase at $T_c = 2/\log(1+\sqrt{2})\approx 2.269815$. 

\begin{figure} 
    \centering
    \includegraphics[width=1\linewidth]{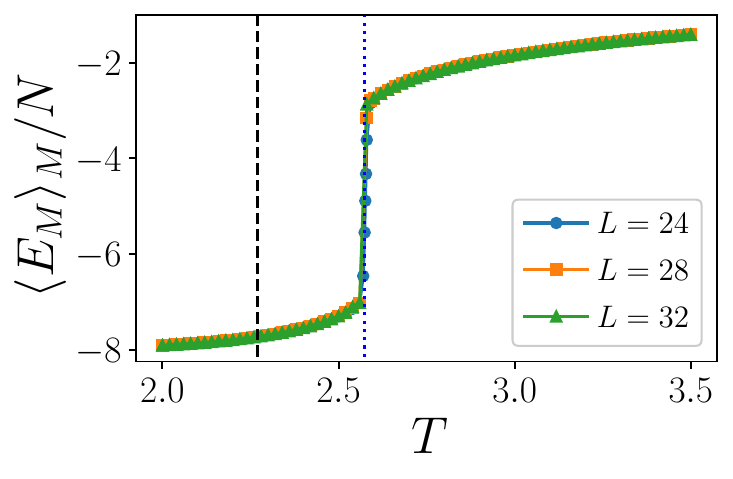}
    \includegraphics[width=1\linewidth]{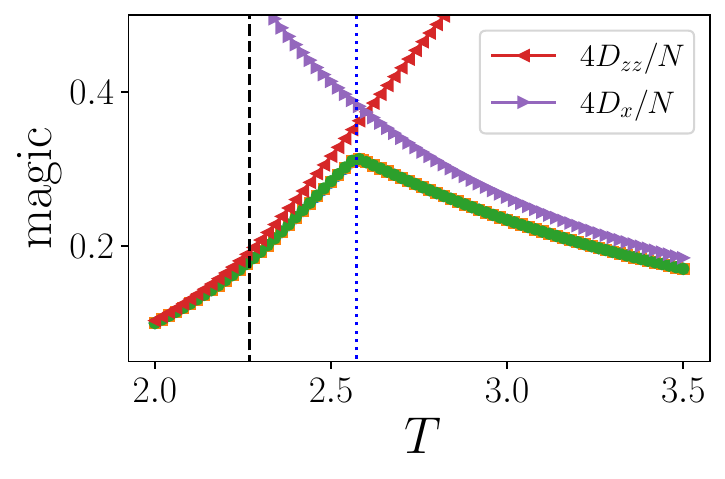}
    \caption{Magic and stabilizer bounds for the SMF 2D Ising ferromagnet. The top panel shows the behaviour of $\langle E_M \rangle_M$ introduced in Sec.~\ref{sec:SMFsre} as a stepping stone to compute $M_2/N$ (bottom panel). The vertical dashed lines indicate the location of the quantum phase transition, whereas the vertical dotted lines indicate the location of the transition in the coupled layered system $Z_M$ (resulting in a cusp in the magic $M_2$).}
    \label{fig:2dising}
\end{figure}

To study the thermodynamics properties of $Z_M$, we perform Monte Carlo simulations augmented with Wolff cluster algorithm~\cite{wolff1989} and parallel tempering~\cite{Hukushima1996,marinari1996}.
We study the energy $\langle E_M \rangle_M$ as a function of temperature for a range of system sizes as shown in Fig.~\ref{fig:2dising} (top panel). We observe a behaviour compatible with a first-order transition from a high-temperature paramagnetic phase to a low-temperature ordered phase, at some value $T^* \neq T_c$ (whereas we know $\langle E \rangle$ from the classical 2D Ising model to be smooth, with a singularity in its first derivative at $T_c$). 
The presence of a first order transition in the classical system described by $Z_M$ will be a common feature in all examples considered in our work; however, while $T^* > T_c$ for the 2D Ising case, we will find for example that $T^* < T_c$ in higher-dimensional systems. 

Integrating the energy as in Eq.~\eqref{eq:m2_derivative} from high temperature, we obtain the SRE $M_2$. In contrast to the results of previous studies~\cite{haug2023quantifying,tarabunga2023manybody}, we see that $M_2$ is continuous and does not exhibit a maximum nor minimum at the transition point. In fact, we know that at $T_c$, $M_2$ inherits a singularity in its second derivative from the singularity in the first derivative of the energy $\langle E \rangle$ of the associated classical system described by the partition function $Z$ (which undergoes a second order phase transition). 
The maximum of $M_2$ occurs instead away from the quantum critical point (into the paramagnetic phase), at a cusp that can be related to the first order transition point of the classical system described by $Z_M$. Furthermore, we observe that the bounds $4D_x$ and $4D_{zz}$ lie very close above $M_2$. By Eq.~\eqref{eq:ineq_dmin2}, this also establishes strict upper and lower bounds for $D_{\text{min}}$. This is again a common feature that we consistently observe in all examples considered in this work. 

\begin{figure} 
    \centering
    \includegraphics[width=1\linewidth]{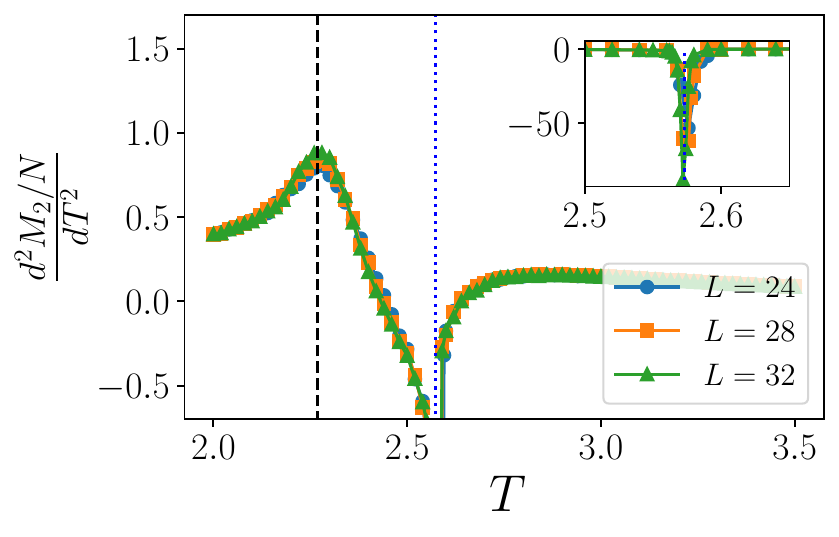}
    \caption{Second derivative of $M_2$ for the 2D SMF Ising ferromagnet. The vertical dashed line indicates the location of the quantum phase transition, whereas the vertical dotted line indicates the location of the transition $T^*$ in the coupled layered system $Z_M$. The inset shows the full extent of $d^2(M_2/N)/dT^2$ near $T^*$, which is truncated in the main plot for visualisation purposes.}
    \label{fig:2dising_d2m}
\end{figure}

Although the SRE $M_2$ appears relatively featureless across $T_c$, we know from Eq.~\eqref{eq:m2_derivative} that it must inherit any singularity present in $\langle E \rangle$ and in $\langle E_M \rangle_M$. The well-known critical behaviour of the 2D Ising model gives a singularity in the second derivative of $M_2$, which is related to the specific heat of the classical 2D Ising model: $d^2M_2/dT^2$ exhibits a peak at $T_c$ which diverges logarithmically, as shown in Fig.~\ref{fig:2dising_d2m}. A (negative) peak is observed at $T^*$, due to the the specific heat of $Z_M$; here the transition is first order and the peak diverges as $N$, 
much faster than the known logarithmic divergence at $T_c$.

We note that, by Wegner duality~\cite{Wegner}, the SMF groundstates corresponding to the 2D Ising model are dual to a wavefunction deformation of the toric code studied in Ref.~\cite{Castelnovo2008}. Therefore, the SREs of the two wavefunctions are identical (up to a constant shift), since the SREs are preserved by Wegner duality~\cite{tarabunga2023manybody}.
%
%

\subsection{3D SMF Ising ferromagnet}
The discussion of the 3D Ising ferromagnet goes along the same line as in 2D, with the spins $\sigma_i$ living without loss of generality on the 3D cubic lattice. The model is known to exhibit a second-order phase transition between the ferromagnetic and the paramagnetic phase. Through large-scale Monte Carlo simulations, the critical point was found to be at $T_c\approx4.5115$~\cite{Talapov1996}.

\begin{figure} 
    \centering
    \includegraphics[width=1\linewidth]{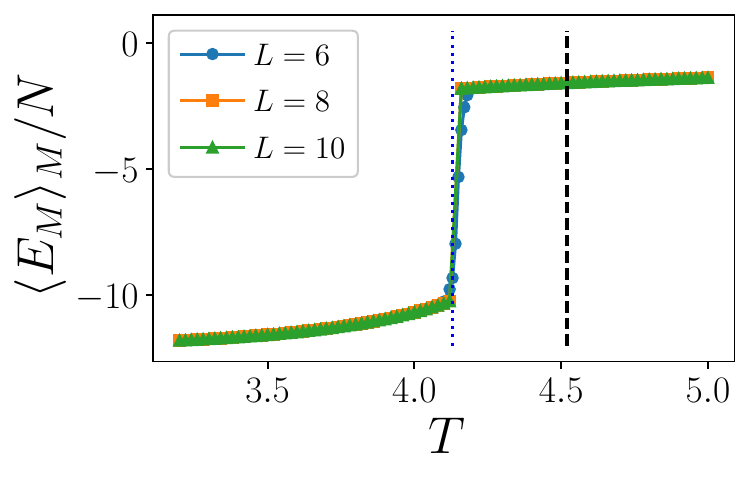}
    \includegraphics[width=1\linewidth]{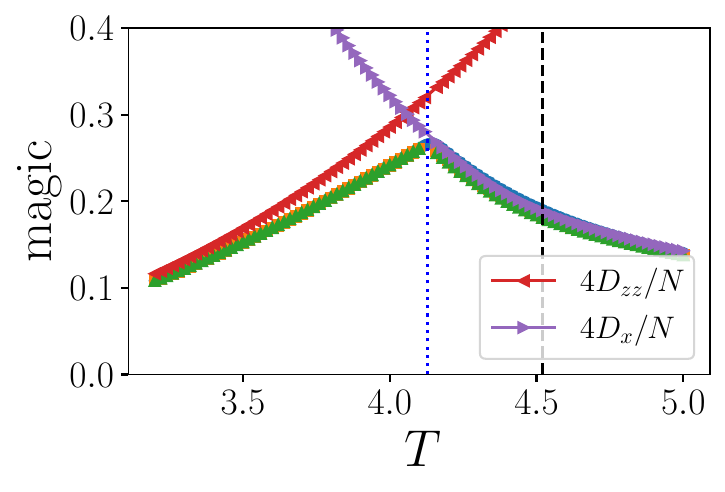}
    \caption{Magic and stabilizer bounds for the SMF 3D Ising ferromagnet. The top panel shows the behaviour of $\langle E_M \rangle_M$, and the bottom panel shows the SRE density $M_2/N$ (same system sizes). The vertical dashed lines indicate the location of the quantum phase transition, whereas the vertical dotted lines indicate the location of the transition in the coupled layered system $Z_M$ (resulting in a cusp in the magic $M_2$).}
    \label{fig:3dising}
\end{figure}

The results are shown in Fig.~\ref{fig:3dising}. The energy $\langle E_M \rangle_M$ once again exhibits a first-order transition that induces a cusp in the behaviour of $M_2$ at $T^*$, where it reaches its maximum value. At the quantum phase transition, $M_2$ is once again continuous, with a singularity in its second derivative. Differently from the 2D case, the cusp (and maximum) of $M_2$ occurs in the ferromagnetic phase instead of the paramagnetic one. 
Once again, the upper bounds $4D_x$ and $4D_{zz}$ lie very close to $M_2$. 
%
%

\subsection{Infinite-range Ising model}
For completeness, we consider the case of an infinite-range Ising ferromagnet, with classical energy 
\begin{align}
    E_\sigma &= -\frac{1}{2N}\sum_{i\neq j} \sigma_i \sigma_j \\
    &= -\frac{1}{2N} \left( \sum_{i} \sigma_i \right)^2 + 1/2 
    \, , 
\end{align}
which becomes again analytically tractable. Hereafter, we shall neglect the trivial constant energy shift in the last line. 

The partition function of the infinite-range model can be evaluated by first recasting it to a Gaussian integral and then performing a saddle-point approximation, which is exact in the thermodynamic limit $L\to \infty$ (see e.g., Ref.~\cite{nishimori2001}). Explicitly, the free energy is given by
\begin{equation} \label{eq:free_energy_infinite}
    \beta F/N = \frac{\beta}{2} m^2 - \log\left[2 \cosh(\beta m)\right]
    \, ,
\end{equation}
where the saddle point magnetisation is found by solving the self-consistency equation 
\begin{equation}
    m = \tanh(\beta m)
    \, .
\end{equation}
The system exhibits a second-order phase transition at $T_c=1$ between the ferromagnetic and the paramagnetic phase.

The evaluation of $\beta F_M=-\log Z_M$ follows along the same lines. First, we write the partition function as
\begin{eqnarray}
    Z_M &=& \sum_{\sigma} \exp\left[\frac{\beta}{4N}\sum_{a=1}^{4}\left(\sum_i \sigma_i^{(a)}\right)^2 
    \right.
    \nonumber \\
    && \qquad \quad \left. 
    + \frac{\beta}{4N}\sum_{a=1}^{4}\left(\sum_i  \prod_{b\neq a} \sigma_i^{(b)}\right)^2\right]
    \, .
\end{eqnarray}
Then, we make use of the identity
\begin{equation}
    e^{\alpha x^2/2} = \sqrt{\frac{\alpha N}{2\pi}} \int_{-\infty}^{\infty} dm \: e^{-N \alpha m^2/2+\sqrt{N}amx} 
\end{equation}
to obtain
\begin{align}
\label{eq:zm_infinite}
    Z_M &= \left(\frac{\beta N}{4\pi}\right)^4 \int_{-\infty}^{\infty} \prod_{a=1}^4 dm_a \prod_{b=1}^4 dq_b 
    \\
    &\times 
    \exp\left[-\frac{N\beta m_a^2}{4}-\frac{N\beta q_b^2}{4}+ N\beta \tilde{F}  \right]
    \, ,
\end{align}
where
\begin{eqnarray}
    e^{\beta \tilde{F}} = \sum_{\eta^{(1,...,4)} = \pm 1} \exp\left[ \sum_a \frac{\beta m_a}{2} \eta^{(a)} 
    + \sum_b \frac{\beta q_b}{2}  \prod_{c\neq b} \eta^{(c)}\right]
    \, . 
    \nonumber 
\end{eqnarray}

In the limit $L\to \infty$, we can evaluate the above integral using the saddle-point approximation, such that the free energy is given by
\begin{equation}
    \beta F_M = \frac{N\beta }{4}  \sum_a m_a^2 +\frac{N \beta}{4} \sum_b q_b^2  - N \beta \Tilde{F}.
\end{equation}
Taking partial derivative with respect to all $m_a$ and $q_b$, we obtain the self-consistent equations
\begin{equation} \label{eq:self-const-m}
    m_a = 
    \frac{\sum \eta^{(a)} \exp\left[\sum_c \frac{ \beta m_c}{2} \eta^{(c)} + \sum_d \frac{ \beta q_d}{2}  \prod_{c\neq d} \eta^{(c)}\right] }
    {\sum \exp\left[\sum_c \frac{ \beta m_c}{2} \eta^{(c)} + \sum_d \frac{ \beta q_d}{2}  \prod_{c\neq d} \eta^{(c)}\right] }
    \, ,
\end{equation}
and
\begin{equation} \label{eq:self-const-q}
    q_b = 
    \frac{\sum \prod_{c\neq b} \eta^{(c)}  \exp\left[\sum_c \frac{ \beta m_c}{2} \eta^{(c)} + \sum_d \frac{ \beta q_d}{2} \prod_{c\neq d} \eta^{(c)}\right] }
    {\sum \exp\left[\sum_c \frac{ \beta m_c}{2} \eta^{(c)} + \sum_d \frac{ \beta q_d}{2}  \prod_{c\neq d} \eta^{(c)}\right] }
    \, ,
\end{equation}
respectively, for $a,b=1,2,3,4$. The outer summations above are over $\eta^{(1)},\eta^{(2)},\eta^{(3)},\eta^{(4)}=\pm1$. The quantity $m_a$ corresponds to the magnetization of the $a$-th layer, 
\begin{equation}
    m_a = \frac{1}{N} \left\langle \sum_i \sigma_i^{(a)}\right\rangle 
    \, . 
\end{equation}
while $q_b$ corresponds to
\begin{equation}
    q_b = \frac{1}{N} \left\langle \sum_i \prod_{c\neq b} \sigma_i^{(c)} \right\rangle 
    \, . 
\end{equation}

The procedure outlined above can be straightforwardly generalized to higher (integer) index $n>2$. If we assume, by symmetry, that the solution satisfies $m_1=m_2=...=m_{2n}=q_1=...=q_{2n}=m$~\footnote{The symmetry between $m_a$ and $q_b$ may not be immediately obvious, but it can be seen as follows: for each site $i$ in the layer $a$, define spin $s_i^{(a)} = \prod_{c\neq a} \sigma_i^{(c)}$. After this change of variable, we obtain $q_a = \frac{1}{N} \left\langle \sum_i s_i^{(a)}\right\rangle 
    \,$ and $m_b = \frac{1}{N} \left\langle \sum_i \prod_{c\neq b} s_i^{(c)} \right\rangle 
    \,$. Namely, the role of $m_a$ and $q_b$ is interchanged after the change of variable.  }, then the self-consistent equations simplify to
\begin{equation} \label{eq:self-const-sym}
    m = \frac{\cosh(\beta m)^{2n-1} \sinh(\beta m) + \sinh(\beta m)^{2n-1} \cosh(\beta m)}{1+ \cosh(\beta m)^{2n} + \sinh(\beta m)^{2n}}
\end{equation}
One can verify that Eq.~\eqref{eq:self-const-sym} always admits $m=0$ as a solution, which minimizes $F_M$ at high temperature; it also admits the solution $m=1$, which minimizes $F_M$ at $T=0$, and the transition between the two is first-order. 

\begin{figure} 
    \centering
    \includegraphics[width=1\linewidth]{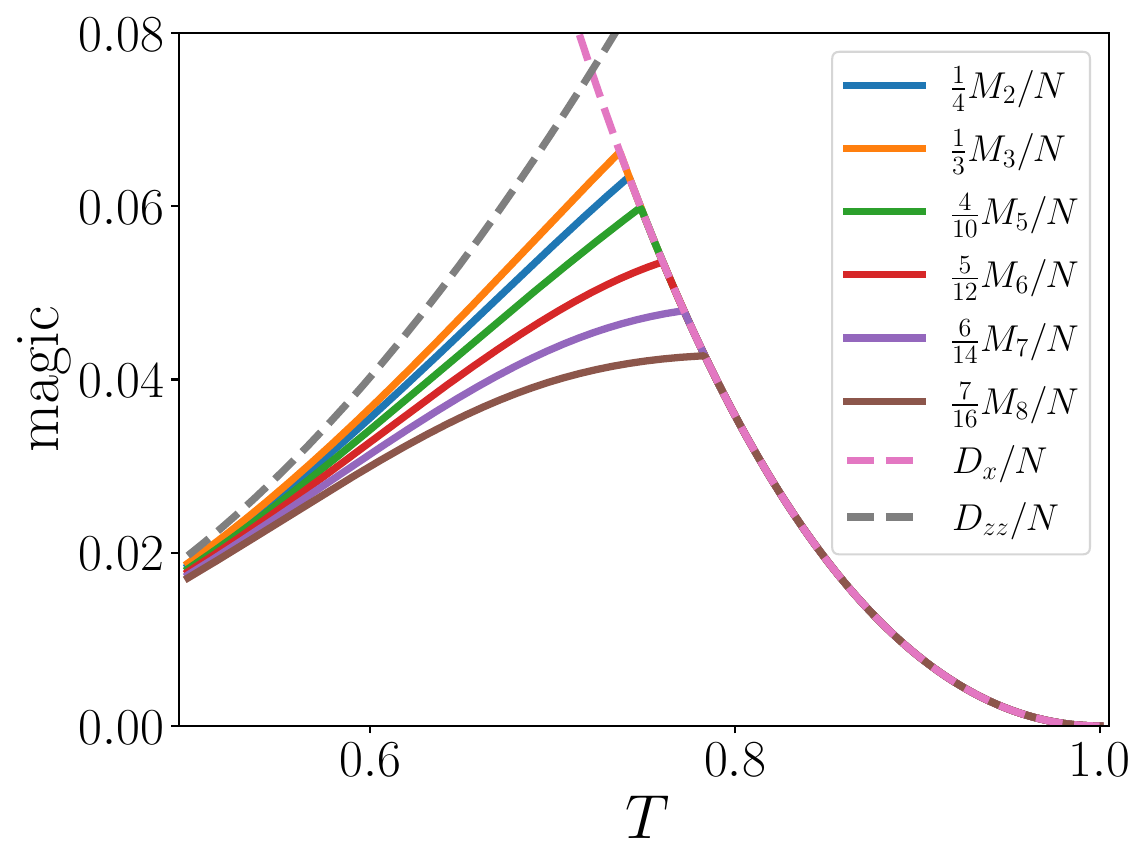}
    \caption{Magic and stabilizer bounds for the SMF infinite-range Ising ferromagnet. For each $n$, we plot $\frac{n-1}{2n} M_n / N$, which are upper bounded by $D_x$ and $D_{zz}$ by Eq. \eqref{eq:ineq_dmin}. The system undergoes a quantum phase transition at $T_c=1$.}
    \label{fig:infinite_range}
\end{figure}

We solve Eq.~\eqref{eq:self-const-sym} numerically and compute $M_n$ for $n\in\{2,3,4,5,6,7,8 \}$ (see Fig.~\ref{fig:infinite_range}). In the limit $n \to \infty$, Eq.~\eqref{eq:self-const-sym} reduces to $m=\tanh(\beta m)$, which is exactly the self-consistent equation for the infinite-range model. This implies that $M_n \to 0$ as $n \to \infty$, as expected.

Furthermore, $D_x$ and $D_{zz}$ can be computed directly from Eq.~\eqref{eq:Dx} and~\eqref{eq:Dzz}, respectively, using the free energy in Eq.~\eqref{eq:free_energy_infinite}. We plot them along the SREs in Fig.~\ref{fig:infinite_range}. 

Similarly to the 3D Ising ferromagnet, and contrary to the 2D case, the cusp (and maximum) of $M_2$ occurs well within the ferromagnetically ordered phase. In fact, the behavior of $M_2$ along with $D_x$ and $D_{zz}$ are very similar to the 3D case. Interestingly, the stabilizer bound to the magic is exactly met by $D_x$ for any $T$ larger than the cusp value. This is because in this regime $\log Z_{M,n} = 2n \log 2$, while $\log Z(\beta/2)= \log 2$ in Eq.~\eqref{eq:Dx}, which implies $ M_n = \frac{2n}{n-1} D_x$. By Eq.~\eqref{eq:ineq_dmin}, it follows that $D_\text{min} = \frac{n-1}{2n} M_n$ for any $T$ larger than the cusp value of an index $n$. For $T \geq 1$, all $M_n$ and $D_x$ vanish, i.e., the states are asymptotically close to the stabilizer state $|+++... \rangle$.
%
%

\subsection{J1-J2 model and first order behaviour}
Up to now we considered quantum SMF Hamiltonians that exhibit continuous phase transitions. Here we investigate what happens at a first order quantum phase transition by looking at the SMF $J_1$-$J_2$ Ising model on the square lattice, related to a classical model with energy 
\begin{equation}
    E_{\sigma} = - J_1 \sum_{\langle ij \rangle} \sigma_i \sigma_j + J_2 \sum_{\langle\langle ij \rangle\rangle} \sigma_i \sigma_j
    \, ,
\end{equation}
where $J_1,J_2>0$. The first term corresponds to a ferromagnetic Ising nearest-neighbour interaction, while the second term is an antiferromagnetic interaction across the diagonals of the square plaquettes. For an appropriate choice of the system parameters, e.g., when the ratio $g=J_1/J_2=0.55$, the system exhibits a first-order transition between a high-temperature paramagnetic phase and a low-temperature stripe phase at $T_c \approx 0.772$ (setting $J_1=1$ as the reference energy scale)~\cite{Sandvik2010}. In the stripe phase, the ground states are fourfold degenerate, and can be understood as two decoupled antiferromagnetic ground states.

\begin{figure} 
    \centering
    \includegraphics[width=1\linewidth]{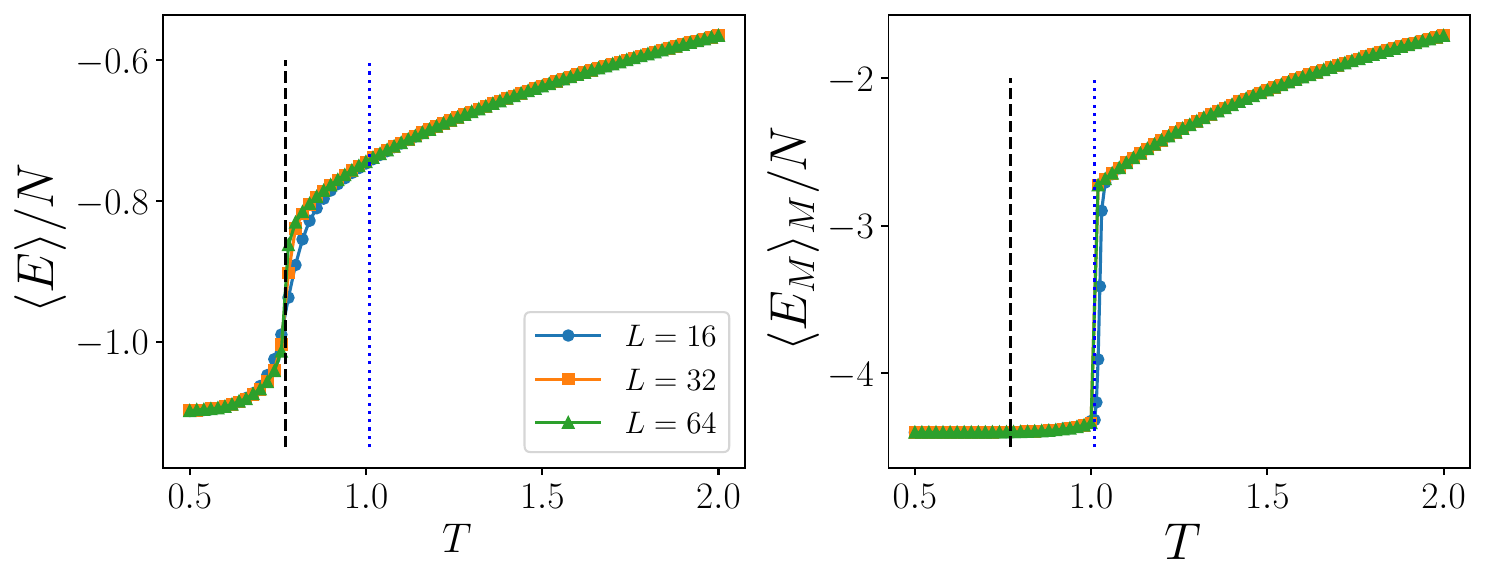}
    \includegraphics[width=1\linewidth]{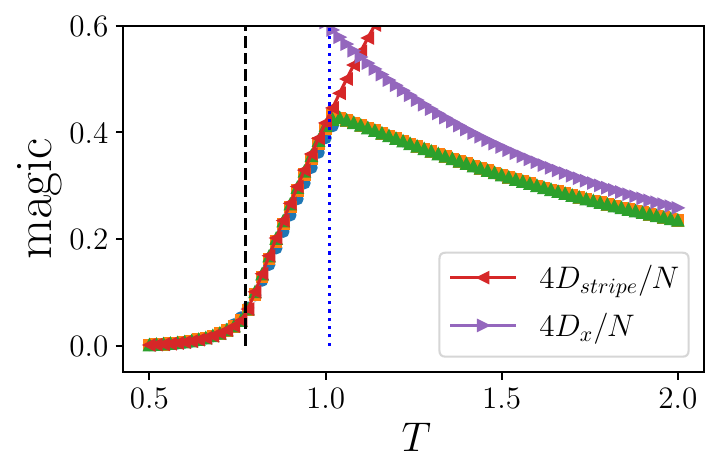}
    \caption{Magic and stabilizer bounds for the SMF 2D $J_1$-$J_2$ Ising model. The top panels show the behaviour of $\langle E \rangle$ and $\langle E_M \rangle_M$, and the bottom panel shows the SRE density $M_2/N$ (same system sizes). The vertical dashed lines indicate the location of the quantum phase transition, whereas the vertical dotted lines indicate the location of the transition in the coupled layered system $Z_M$ (resulting in a cusp in the magic $M_2$).}
    \label{fig:j1-j2}
\end{figure}

We also observe a first order phase transition in the associated coupled layered system $Z_M$, but at a higher temperature $T^*$, well into the paramagnetic phase (as in the 2D Ising ferromagnet, and contrary to 3D and infinite-range). Therefore, in this system we expect two discontinuities in the first derivative of $M_2$ with respect to $T$: One at the quantum phase transition ($T_c$), where the slope is positive on both sides and approximately doubles across it; the other at $T^*$, where the slope changes sign abruptly, giving rise once again to a maximum in $M_2$, where a cusp occurs.

In Fig.~\ref{fig:j1-j2} we also compare $M_2$ with the bounds provided by the paramagnetic ($D_x$, asymptotically accurate for $T \to \infty$) and stripe ($D_{\rm stripe}$, asymptotically accurate for $T \to 0$) phases. The latter appears to be remarkably close for all $T < T^*$. 

%
%

\subsection{Antiferromagnetic triangular Ising model}
We now proceed to a more exotic model where the quantum SMF Hamiltonian is related to the classical antiferromagnetic triangular Ising model~\cite{Wannier1950,Houtappel1950}, with energy
\begin{equation} 
    E_{\sigma} = \sum_{\langle ij \rangle} \sigma_i \sigma_j,
\end{equation}
where the spins $\sigma_i$ live on the sites of a triangular lattice, $i=1,\ldots,N$. The model features an extensive ground state degeneracy with algebraically decaying correlations at $T=0$, while it is disordered at all temperatures $T \neq 0$~\cite{Wannier1950}. 

We first show that the classical system described by $Z_M$ also features an extensive ground state degeneracy. To this end, we provide a lower bound on the zero-point entropy by explicitly constructing an extensive set of states with lowest energy. To do so, let us divide the triangular lattice on each layer into three sublattices. Let us then fix the spins on two of the sublattices to $1$ and $-1$, respectively, with the same choice for all four layers. One can then straightforwardly verify that the spins on the remaining sublattice on each layer can be chosen arbitrarily without affecting the energy $E_M$ of the system, and that the latter is indeed minimised. The number of such states is $2^{4N/3}$, which implies 
\begin{equation}
    S_M(0) \geq \frac{4N}{3} \ln 2 \, .
\end{equation}
As one can straightforwardly think of other configurations that minimize the energy, this bound is not tight. 

\begin{figure} 
    \centering
    \includegraphics[width=1\linewidth]{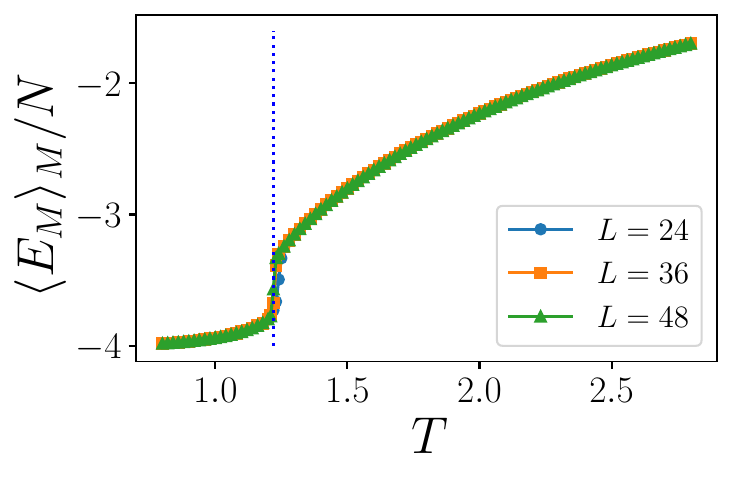}
    \includegraphics[width=1\linewidth]{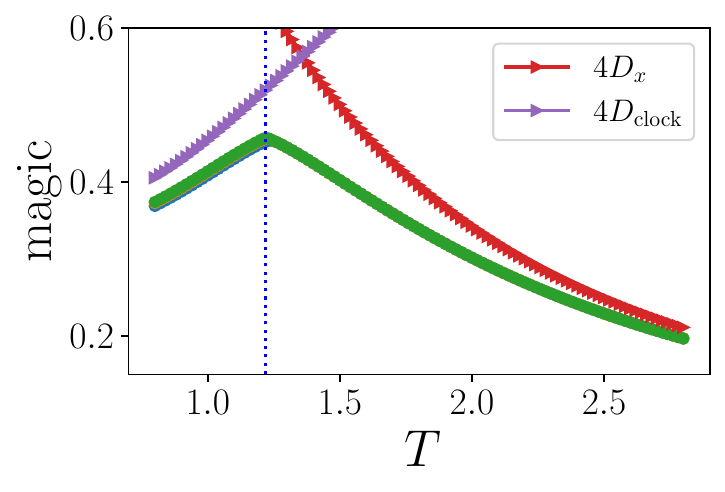}
    \caption{Magic and stabilizer bounds for the SMF triangular antiferromagnetic Ising model. The top panel shows the behaviour of $\langle E_M \rangle_M$, and the bottom panel shows the SRE density $M_2/N$ (same system sizes). The vertical dotted lines indicate the location of the transition in the coupled layered system $Z_M$ (resulting in a cusp in the magic $M_2$).
    }
    \label{fig:triangular_afm}
\end{figure}

We find that the corresponding $Z_M$ features a phase transition that appears to be first-order, as displayed in Fig.~\ref{fig:triangular_afm}, albeit of a less strong nature than in the cases discussed previously. Once again, $M_2$ exhibits a cusp at the transition point $T^*$ of the classical system described by $Z_M$. 

Unlike in the other models considered so far, in this case the ground state at $T=0$ is not a stabilizer state. We note that the configurations with the three sublattice structure given above is also known as the clock state, which arises as the quantum ground state of the quantum triangular Ising antiferromagnet at small magnetic field~\cite{Blankschtein1984, Moessner2001, Isakov2003,Wang2017}. While it is not the exact ground state for $T=0$, it constitutes a significant part of the ground state. Thus, we compare $M_2$ with $D_{\text{clock}}(|\psi \rangle) = D(|\psi \rangle,|\phi \rangle)$ where $|\phi \rangle$ is the clock state defined above, which is a stabilizer state. $D_{\text{clock}}$ is obtained in a similar way as $D_{zz}$ (see Sec. \ref{sec:bound_m2}). At $T=0$, $D_{\text{clock}}$ is given by 
\begin{eqnarray}
    D_{\text{clock}}(T=0) = S(0) - \frac{N}{3} \ln 2
\, ,
\end{eqnarray}
where $S(0) \simeq 0.3383 \, N$ is the zero-point entropy of the antiferromagnetic triangular Ising model~\cite{Wannier1950}. On the other hand, $M_2$ is given by 
\begin{equation}
    M_2(T=0) = 4 S(0) - S_M(0)
\, .
\end{equation}
This is obtained by setting $T=0$ in Eq.~\eqref{eq:sre_smf}. In this case, the observation that $4D_{\text{clock}}$ is strictly larger than $M_2$ can be attributed to the fact that the zero-point entropy $S_M(0)$ is  strictly larger than $\frac{4N}{3} \ln 2$. In turn, this is also a manifestation of the fact that the ground state of the classical system  described by $Z(T)$ is not a stabilizer state for $T=0$.
%
%
%

\subsection{Edwards-Anderson model}
Finally, we consider an example of a disordered system, where the quantum SMF Hamiltonian is related to the Edwards-Anderson (EA) model, with energy
\begin{equation}
    E_{\sigma} = - \sum_{\langle ij \rangle} J_{ij} \sigma_i \sigma_j
    \, ,
\end{equation}
where the spins $\sigma_i$ live on the 3D cubic lattice. Here, the couplings $J_{ij}$ are independently drawn from a  Gaussian distribution with zero mean and unit variance. The system is known to undergo a continuous transition from the high-temperature phase to the low-temperature spin glass phase at $T_c \approx 0.95$~\cite{Katzgraber2006}.

This model has a unique ground state for any realization of $J_{ij}$ (up to global spin flip). For system sizes up to $L=10$, the exact ground states and their energy can be readily obtained using the McGroundstate server~\cite{CJMM22}.

\begin{figure} 
    \centering
    \includegraphics[width=1\linewidth]{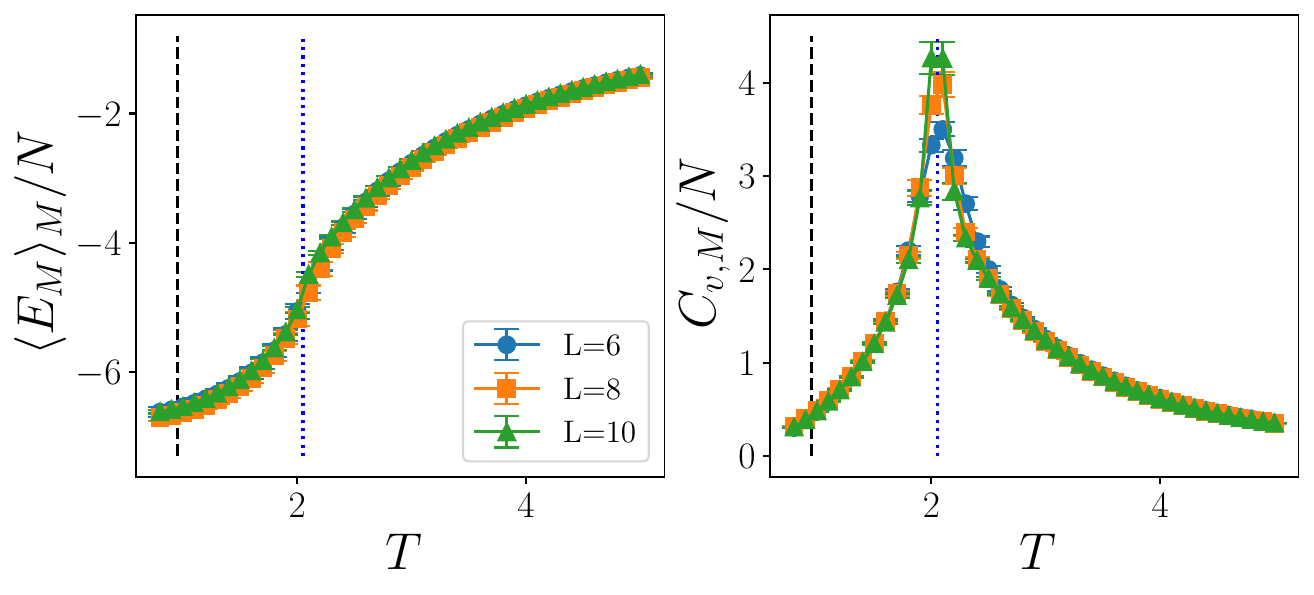}
    \includegraphics[width=1\linewidth]{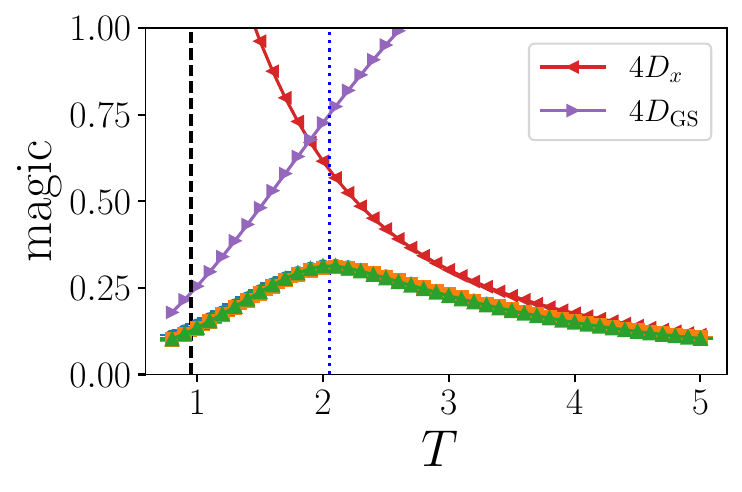}
    \caption{Magic and stabilizer bounds for the 3D SMF EA model. The top panels show the behaviour of $\langle E_M \rangle_M$ (left) and of the specific heat $C_{v,M}$ (right), and the bottom panel shows the SRE density $M_2/N$ (all for the same set of system sizes). The vertical dashed lines indicate the location of the quantum phase transition, whereas the vertical dotted lines indicate the location of the transition at $T^*$ in the coupled layered system $Z_M$ (resulting in a maximum in the magic $M_2$). The results are averaged over $100$, $50$, and $30$ realizations for linear sizes $L=6,8,10$, respectively.}
    \label{fig:edwards_anderson}
\end{figure}

\begin{figure} 
    \centering
    \includegraphics[width=1\linewidth]{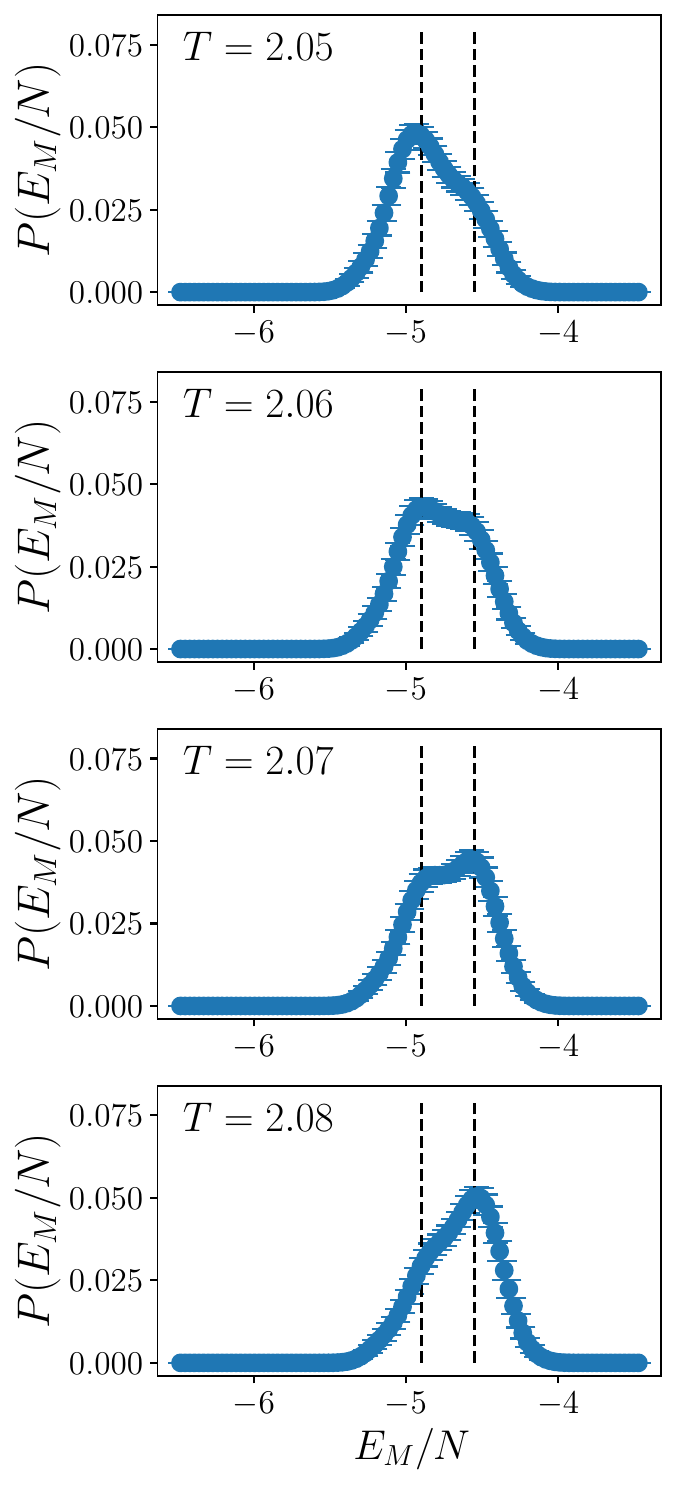}
    \caption{Energy histograms of the coupled layered system $Z_M$ corresponding to the 3D EA model, in thermodynamic equilibrium at different temperatures for a system of size $L=12$. The vertical dashed lines are guides to the eye tracking the (same) location location of the two peaks across the panels.}
    \label{fig:edwards_anderson_histogram}
\end{figure}

We show the energy $\langle E_M \rangle_M$, the specific heat $C_{v,M}$ and the magic $M_2$ in Fig.~\ref{fig:edwards_anderson}. Again, the maximum occurs well into the paramagnetic phase, at the transition point of the coupled layered system $Z_M$. 
Unlike in the previous cases, where the first order behaviour of the $Z_M$ transition was self-evident because of the noticeable discontinuity in $\langle E_M \rangle_M$, the situation is less clear-cut here. 
While the maximum of the specific heat appears to grow slower than $N$, suggesting a second order transition, we are unable to identify a clear scaling of the specific heat within the accessible system sizes.
We further compute equilibrium energy histograms at different temperatures around $T^*$, shown in Fig.~\ref{fig:edwards_anderson_histogram}. 
The behaviour closely resembles a trade off between two different peaks, whose positions are approximately temperature-independent (although we are unable to see the minimum in between them scale to zero as a function of system size, within the systems accessed in this work). Overall, we suggest that the transition in this model is weakly first-order. 

In Fig.~\ref{fig:edwards_anderson} we also compare $M_2$ with bounds from $D_x$ and $D_{\text{GS}}$, where $D_{\text{GS}}$ is obtained from the overlaps with the exact ground state for each realization, computed by thermodynamic integration as discussed in Sec.~\ref{sec:bound_m2}, using the McGroundstate server~\cite{CJMM22} to obtain the exact ground state energy. In this case, the bounds are somewhat higher than encountered in previous cases. Nevertheless, the crossing between $D_x$ and $D_{\text{GS}}$ still occurs close to the maximum of $M_2$. 

\begin{figure} 
    \centering
    \includegraphics[width=1\linewidth]{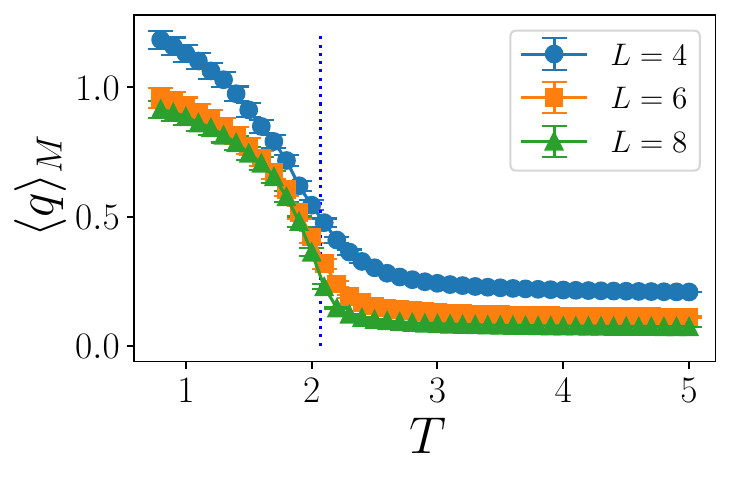}
    \caption{Spin overlap $\langle q \rangle_M$ for the coupled layered system $Z_M$ corresponding to the 3D EA model, for system sizes $L\in \{4,6,8\}$.}
    \label{fig:edwards_anderson_rsb}
\end{figure}

Finally, we investigate the nature of the low-temperature phase of $Z_M$. A natural candidate is a spin glass phase, accompanied by replica symmetry breaking (RSB), akin to the low-temperature phase of the 3D EA model. To detect RSB, we compute the spin overlap
\begin{equation}
    q = \frac{1}{N} \sum_{i=1}^N \langle \sigma_i^\alpha \sigma_i^\beta \rangle
    \, ,
\end{equation}
where $\alpha$ and $\beta$ represents two copies of the system with the same disorder. We show the spin overlap in the coupled system $Z_M$ in Fig.~\ref{fig:edwards_anderson_rsb}. 
It can be seen that the spin overlap is vanishing in the high-temperature paramagnetic phase, while it becomes non-zero in the low-temperature phase, signifying RSB.
%
%

\section{Conclusions
\label{sec:conclusions}}
We introduced a way to compute the SRE~\cite{leone2022stabilizer} with integer Renyi index $n>1$ in terms of wavefunction coefficients in many body systems, that make it amenable to efficient computation using Monte Carlo sampling. We applied this approach to generalized Rokhsar-Kivelson systems whose Hamiltonians allow a stochastic matrix form decomposition~\cite{Castelnovo2005}. Thanks to the known correspondence between ground states of these systems and associated classical statistical mechanics problems, we have been able to express the SRE in terms of classical free energy differences, which can be efficiently computed by thermodynamic integration. Crucially, temperature plays the role of a tunable parameter in the quantum Hamiltonians, allowing us to drive these systems across quantum phase transitions and study the behaviour of their SRE. With this approach we were able to study the SRE of large high-dimensional systems, unattainable using existing tensor network-based techniques, and in some cases obtaining explicit analytical results. 

We applied this insight to a range of quantum many body SMF Hamiltonians, encompassing the Ising ferromagnet in 1D, 2D, 3D, and infinite dimensions; the $J_1-J_2$ model on the square lattice (exhibiting a first order transition); the triangular Ising antiferromagnet (fully frustrated, devoid of ordering); and the Edwards-Anderson model on the cubic lattice (which undergoes a glass transition). Generally, we observed that the behaviour of the SRE is relatively featureless across quantum phase transitions in these systems, although it is indeed singular in its first or higher order derivative, depending on the first or higher order nature of the transition. We found that the maximum of the SRE generically occurs at a cusp away from the quantum critical point, where the derivative suddenly changes sign. 
Curiously, the cusp appears to occur in the disordered phase in two dimensions, and in the ordered phase in higher dimensions, suggesting that it may be altogether unrelated to the ordering behaviour of the quantum system. 

%

We further compared the SRE to the logarithm of overlaps with specific stabilizer states, that are asymptotically realised in the ground state phase diagrams of these systems. We find that they display strikingly similar behaviors, which in turn establish rigorous bounds on the min-relative entropy of magic. 

In our work we were able to make some progress in understanding the behaviour of the magic and its maximum in many body quantum (SMF) systems, throughout their phase diagrams, in terms of partition functions and thermodynamic properties of associated classical problems, and by comparing it with overlaps of asymptotic stabilizer states. One wonders whether further progress could be made using field theoretic approaches for the associated classical problems, in particular $\varepsilon$-expansions just above 2D or just below 3D to shed light on the location of the SRE maximum with respect to the quantum phase transitions. We shall leave these and other interesting open questions for future work. 

As we discussed at the end of Sec.~\ref{sec:Ising2D}, our results for the 2D SMF Ising ferromagnet straightforwardly extend to the toric code~\cite{Kitaev2006} and SMF variations thereof~\cite{Castelnovo2008}, which are some of the simplest examples of $\mathbb{Z}_2$ lattice gauge theories. It will be interesting to consider other quantum SMF Hamiltonians constructed from classical systems that exhibit an emergent gauge symmetry, such as dimer~\cite{Castelnovo2007,Moessner2003}, (spin) ice~\cite{Hermele2004,Castro2006,Benton2012} and vertex models in general. Simulating these systems in their low temperature phases typically requires the use of loop updates, which pose a nontrivial challenge for the partition function $Z_M$ introduced in Sec.~\ref{sec:SMFsre}, and is beyond the scope of the present work. 
%
%

\section*{Acknowledgements}
We are grateful to Marcello Dalmonte, Emanuele Tirrito, Simon Trebst and Guo-Yi Zhu for useful discussions, and to Jorge Kurchan for suggesting the $\varepsilon$-expansion direction for future work. 
P.S.T. acknowledges support from the Simons Foundation through Award 284558FY19 to the ICTP.
This work was partly supported by the PNRR MUR project PE0000023-NQSTI, and by the EU-Flagship programme Pasquans2.
For the purpose of open access, the authors have applied a Creative Commons Attribution (CC BY) licence to any Author Accepted Manuscript version arising from this submission. 
This research was funded in part by the Engineering and Physical Sciences (EPSRC) grant No. EP/M007065/1 and EP/V062654/1. 
%
%
\appendix

\section{Derivation of the expression for $M_2$ in Sec.~\ref{sec:SMFsre}
\label{sec:proof}}
Consider the single-qubit Pauli operators (including the identity) $P\in \{ I,X,Y,Z \}$. It is convenient to label them using a pair of indices $a,a'\in \{ 0,1\}$, such that $P_{a,a'} = i^{aa'} X^a Z^{a'}$ whereby $P_{0,0} = I$, $P_{1,0} = X$, $P_{0,1} = Z$, and $P_{1,1} = Y$. For a system of $N$ qubits, the Pauli strings can then be written as 
\begin{equation}
    \begin{split}
    P_{\mathbf{a},\mathbf{a'}} =P_{a_1,a_1'} P_{a_2,a_2'} ...  P_{a_N,a_N'} \, ,
    \end{split}
\end{equation}
where $\mathbf{a},\mathbf{a'}$ are $n$-bit strings. We also denote $s = (-1)^a$, for later convenience.

Let the group of all $N-$qubit Pauli strings be $\mathcal{P}_N$. The SRE $M_2$ is defined as:
\begin{equation} \label{eq:M2_def}
\begin{split}
\exp(-M_{2}) &=  \frac{1}{d} \sum_{P \in \mathcal{P}_n} \langle \psi | P | \psi \rangle^{4}  
    = \frac{1}{d}  \sum_{\mathbf{a},\mathbf{a'}} \langle \psi | P_{\mathbf{a},\mathbf{a'}} | \psi \rangle^{4},
\end{split}
\end{equation}
where $d=2^N$ is the Hilbert space dimension.
Substituting $| \psi \rangle = \sum_{{\sigma}} c_{\sigma} | \sigma \rangle $, and using the representation of $P_{a,a'}$ above, we obtain 
\begin{widetext}
\begin{equation}
\begin{split}
\exp(-M_{2}) 
&= 
\frac{1}{d} \sum_{\mathbf{a},\mathbf{a'}} 
\left[ 
\sum_{\sigma,\sigma'} c_{\sigma} c^*_{\sigma'} \langle \sigma' | P_{\mathbf{a},\mathbf{a'}} | \sigma \rangle 
\right]^{4} 
\\
&= \frac{1}{d} \sum_{\mathbf{a},\mathbf{a'}} 
\left[ 
\sum_{\sigma,\sigma'} c_{\sigma} c^*_{\sigma'} 
\prod_{j=1}^N 
i^{a_j a'_j} \langle \sigma'_j | X^{a_j} Z^{a'_j} | \sigma_j \rangle 
\right]^{4} 
\\
&= \frac{1}{d} \sum_{\mathbf{a},\mathbf{a'}} 
\left[ 
\sum_{\sigma,\sigma'} c_{\sigma} c^*_{\sigma'} 
\prod_{j=1}^N 
 \sigma_j^{a'_j} \langle \sigma'_j | X^{a_j} | \sigma_j \rangle 
\right]^{4} 
\\
&= \frac{1}{d} \sum_{\mathbf{a},\mathbf{a'}} 
\left[ 
\sum_{\sigma,\sigma'} c_{\sigma} c^*_{\sigma'} 
\prod_{j=1}^N 
\sigma_j^{a'_j} \langle \sigma'_j | s_j \sigma_j \rangle 
\right]^{4} 
\\
&= \frac{1}{d } \sum_{\mathbf{a},\mathbf{a'}} 
\left[ 
\sum_{\sigma} c_{\sigma} c^*_{s\sigma} 
\prod_{j=1}^N 
\sigma_j^{a'_j} 
\right]^{4} 
\, , 
\end{split}
\end{equation}
where $s\sigma$ represents the tensor product state label where $\sigma'_j = s_j \sigma_j$, $\forall\,j$. Note that the factor $\prod_j i^{a_j a'_j}$ in the second line is independent of $\sigma,\sigma'$, and therefore factors out of the summation and disappears due to the fourth power ($i^4=1$).
Expanding the fourth power explicitly, we can further simplify the expression 
\begin{equation}
\begin{split}
\exp(-M_{2}) &= 
\frac{1}{d} \sum_{\mathbf{a},\mathbf{a'}} 
\sum_{\sigma^{(1)},\sigma^{(2)},\sigma^{(3)},\sigma^{(4)}} 
\left[ 
\prod_{i=1}^4 
c_{\sigma^{(i)}} c^{*}_{s\sigma^{(i)}} 
\prod_{j=1}^N 
\left(\sigma^{(i)}_j\right)^{a'_j} 
\right] 
\\
&= \frac{1}{d} \sum_{\mathbf{a}} 
\sum_{\sigma^{(1)},\sigma^{(2)},\sigma^{(3)},\sigma^{(4)}} 
\left[ 
\prod_{i=1}^4 
c_{\sigma^{(i)}} c^{*}_{s\sigma^{(i)}} 
\right] 
\left[
\prod_{j=1}^N \sum_{a'_j=0,1} 
\left(\prod_{i=1}^4 \sigma^{(i)}_j\right)^{a'_j} 
\right] 
\\ 
&= \frac{1}{d } \sum_{\mathbf{a}} 
\sum_{\sigma^{(1)},\sigma^{(2)},\sigma^{(3)},\sigma^{(4)}} 
\left[ 
c_{\sigma^{(1)}}  c_{\sigma^{(2)}} c_{\sigma^{(3)}}  c_{\sigma^{(4)}} \vphantom{\prod_l\left(1+\sigma_l^{(1)}\sigma_l^{(2)}\sigma_l^{(3)}\sigma_l^{(4)} \right)} 
c^{*}_{s\sigma^{(1)}} c^{*}_{s\sigma^{(2)}} c^{*}_{s\sigma^{(3)}} c^{*}_{s\sigma^{(4)}} \prod_{j=1}^N\left(1+\sigma_j^{(1)}\sigma_j^{(2)}\sigma_j^{(3)}\sigma_j^{(4)}  \right) 
\right]
\, .  
\end{split}
\end{equation}
\end{widetext}

Since the product $\sigma_j^{(1)}\sigma_j^{(2)}\sigma_j^{(3)}\sigma_j^{(4)} \in \{-1,1\}$, the term in square brackets in the last line above is nonzero only if $\sigma_j^{(1)}\sigma_j^{(2)}\sigma_j^{(3)}\sigma_j^{(4)} = 1$ for all sites $j$. This effectively constraints the fourth layer $\sigma^{(4)}=\sigma^{(1)}\sigma^{(2)}\sigma^{(3)}$: 
\begin{eqnarray}
\exp(-M_2) &=& 
\!\!\!\!\sum_{\mathbf{a},\sigma^{(1)},\sigma^{(2)},\sigma^{(3)}} \!
\left[ \vphantom{\sum} 
c_{\sigma^{(1)}}  c_{\sigma^{(2)}} c_{\sigma^{(3)}}  c_{\sigma^{(1)}\sigma^{(2)}\sigma^{(3)}}
\right.
\nonumber \\
&& \qquad \qquad \;\;
\left. \vphantom{\sum}
c^{*}_{s\sigma^{(1)}} c^{*}_{s\sigma^{(2)}} c^{*}_{s\sigma^{(3)}} c^{*}_{s\sigma^{(1)}\sigma^{(2)}\sigma^{(3)}}
\right] 
\, .
\nonumber \\ 
\end{eqnarray}

Finally, we can replace the summation over $\mathbf{a}$ for a summation over $\sigma^{(4)} = s\sigma^{(1)}\sigma^{(2)}\sigma^{(3)}$, to bring the expression into a more symmetric form, 
\begin{eqnarray} 
\exp(-M_2) &=& 
\!\!\!\!\sum_{\sigma^{(1)},\sigma^{(2)},\sigma^{(3)},\sigma^{(4)}}\!
\left[ \vphantom{\sum}
c_{\sigma^{(1)}} c_{\sigma^{(2)}} c_{\sigma^{(3)}}  c_{\sigma^{(1)}\sigma^{(2)}\sigma^{(3)}} 
\right.
\nonumber \\
&& \quad 
\left. \vphantom{\sum}
c^{*}_{\sigma^{(1)} \sigma^{(2)} \sigma^{(4)}} c^{*}_{\sigma^{(1)} \sigma^{(3)} \sigma^{(4)}} c^{*}_{\sigma^{(2)}\sigma^{(3)}\sigma^{(4)}} c^{*}_{\sigma^{(4)}} 
\right] 
\, . 
\nonumber \\ 
\end{eqnarray}
~
%
%

\bibliographystyle{quantum}
\bibliography{bibliography}

\begin{thebibliography}{10}

\bibitem{gottesman1997stabilizer}
Daniel Gottesman.
\newblock ``Stabilizer codes and quantum error correction''~(1997).
\newblock
  \href{http://arxiv.org/abs/quant-ph/9705052}{arXiv:quant-ph/9705052}.

\bibitem{nielsen2002quantum}
Michael~A. Nielsen and Isaac~L. Chuang.
\newblock ``Quantum computation and quantum information''.
\newblock \href{https://dx.doi.org/10.1017/cbo9780511976667}{Cambridge
  University Press}. ~(2012).

\bibitem{vedral1997quantifying}
V.~Vedral, M.~B. Plenio, M.~A. Rippin, and P.~L. Knight.
\newblock ``Quantifying entanglement''.
\newblock \href{https://dx.doi.org/10.1103/PhysRevLett.78.2275}{Phys. Rev.
  Lett. {\bf 78}, 2275--2279}~(1997).

\bibitem{horodecki2009quantum}
Ryszard Horodecki, Pawe\l{} Horodecki, Micha\l{} Horodecki, and Karol
  Horodecki.
\newblock ``Quantum entanglement''.
\newblock \href{https://dx.doi.org/10.1103/RevModPhys.81.865}{Rev. Mod. Phys.
  {\bf 81}, 865--942}~(2009).

\bibitem{smith2006typical}
Graeme Smith and Debbie Leung.
\newblock ``Typical entanglement of stabilizer states''.
\newblock \href{https://dx.doi.org/10.1103/PhysRevA.74.062314}{Phys. Rev. A
  {\bf 74}, 062314}~(2006).

\bibitem{gutschow2010entanglement}
J.~G\"{u}tschow.
\newblock ``Entanglement generation of clifford quantum cellular automata''.
\newblock \href{https://dx.doi.org/10.1007/s00340-009-3840-1}{Applied Physics B
  {\bf 98}, 623--633}~(2009).

\bibitem{preskill2012quantum}
John Preskill.
\newblock ``Quantum computing and the entanglement frontier''~(2012)
  \href{http://arxiv.org/abs/1203.5813}{arXiv:1203.5813}.

\bibitem{harrow2017quantum}
Aram~W. Harrow and Ashley Montanaro.
\newblock ``Quantum computational supremacy''.
\newblock \href{https://dx.doi.org/10.1038/nature23458}{Nature {\bf 549},
  203--209}~(2017).

\bibitem{gottesman1998heisenberg}
Daniel Gottesman.
\newblock ``The heisenberg representation of quantum computers''~(1998).
\newblock
  \href{http://arxiv.org/abs/quant-ph/9807006}{arXiv:quant-ph/9807006}.

\bibitem{gottesman1998theory}
Daniel Gottesman.
\newblock ``Theory of fault-tolerant quantum computation''.
\newblock \href{https://dx.doi.org/10.1103/PhysRevA.57.127}{Phys. Rev. A {\bf
  57}, 127--137}~(1998).

\bibitem{aaronson2004improved}
Scott Aaronson and Daniel Gottesman.
\newblock ``Improved simulation of stabilizer circuits''.
\newblock \href{https://dx.doi.org/10.1103/PhysRevA.70.052328}{Phys. Rev. A
  {\bf 70}, 052328}~(2004).

\bibitem{bravyi2005UniversalQuantumComputation}
Sergey Bravyi and Alexei Kitaev.
\newblock ``{Universal quantum computation with ideal Clifford gates and noisy
  ancillas}''.
\newblock \href{https://dx.doi.org/10.1103/PhysRevA.71.022316}{Phys. Rev. A
  {\bf 71}, 022316}~(2005).

\bibitem{campbell2017roads}
Earl~T. Campbell, Barbara~M. Terhal, and Christophe Vuillot.
\newblock ``Roads towards fault-tolerant universal quantum computation''.
\newblock \href{https://dx.doi.org/10.1038/nature23460}{Nature {\bf 549},
  172--179}~(2017).

\bibitem{bravyi2012magic}
Sergey Bravyi and Jeongwan Haah.
\newblock ``Magic-state distillation with low overhead''.
\newblock \href{https://dx.doi.org/10.1103/PhysRevA.86.052329}{Phys. Rev. A
  {\bf 86}, 052329}~(2012).

\bibitem{chitambar2019}
Eric Chitambar and Gilad Gour.
\newblock ``Quantum resource theories''.
\newblock \href{https://dx.doi.org/10.1103/RevModPhys.91.025001}{Rev. Mod.
  Phys. {\bf 91}, 025001}~(2019).

\bibitem{bravy2016improved}
Sergey Bravyi and David Gosset.
\newblock ``Improved classical simulation of quantum circuits dominated by
  clifford gates''.
\newblock \href{https://dx.doi.org/10.1103/PhysRevLett.116.250501}{Phys. Rev.
  Lett. {\bf 116}, 250501}~(2016).

\bibitem{bravy2016trading}
Sergey Bravyi, Graeme Smith, and John~A. Smolin.
\newblock ``Trading classical and quantum computational resources''.
\newblock \href{https://dx.doi.org/10.1103/PhysRevX.6.021043}{Phys. Rev. X {\bf
  6}, 021043}~(2016).

\bibitem{Howard2017}
Mark Howard and Earl Campbell.
\newblock ``Application of a resource theory for magic states to fault-tolerant
  quantum computing''.
\newblock \href{https://dx.doi.org/10.1103/physrevlett.118.090501}{Physical
  Review Letters{\bf 118}}~(2017).

\bibitem{Heinrich2019}
Markus Heinrich and David Gross.
\newblock ``Robustness of magic and symmetries of the stabiliser polytope''.
\newblock \href{https://dx.doi.org/10.22331/q-2019-04-08-132}{Quantum {\bf 3},
  132}~(2019).

\bibitem{Seddon2021}
James~R. Seddon, Bartosz Regula, Hakop Pashayan, Yingkai Ouyang, and Earl~T.
  Campbell.
\newblock ``Quantifying quantum speedups: Improved classical simulation from
  tighter magic monotones''.
\newblock \href{https://dx.doi.org/10.1103/PRXQuantum.2.010345}{PRX Quantum
  {\bf 2}, 010345}~(2021).

\bibitem{hamaguchi2023handbook}
Hiroki Hamaguchi, Kou Hamada, and Nobuyuki Yoshioka.
\newblock ``Handbook for efficiently quantifying robustness of magic''~(2023).
\newblock  \href{http://arxiv.org/abs/2311.01362}{arXiv:2311.01362}.

\bibitem{leone2022stabilizer}
Lorenzo Leone, Salvatore F.~E. Oliviero, and Alioscia Hamma.
\newblock ``Stabilizer r\'enyi entropy''.
\newblock \href{https://dx.doi.org/10.1103/PhysRevLett.128.050402}{Phys. Rev.
  Lett. {\bf 128}, 050402}~(2022).

\bibitem{liu2022}
Zi-Wen Liu and Andreas Winter.
\newblock ``Many-body quantum magic''.
\newblock \href{https://dx.doi.org/10.1103/PRXQuantum.3.020333}{PRX Quantum
  {\bf 3}, 020333}~(2022).

\bibitem{white2021}
Christopher~David White, ChunJun Cao, and Brian Swingle.
\newblock ``Conformal field theories are magical''.
\newblock \href{https://dx.doi.org/10.1103/PhysRevB.103.075145}{Phys. Rev. B
  {\bf 103}, 075145}~(2021).

\bibitem{Sarkar2020}
S~Sarkar, C~Mukhopadhyay, and A~Bayat.
\newblock ``Characterization of an operational quantum resource in a critical
  many-body system''.
\newblock \href{https://dx.doi.org/10.1088/1367-2630/aba919}{New Journal of
  Physics {\bf 22}, 083077}~(2020).

\bibitem{oliviero2022ising}
Salvatore F.~E. Oliviero, Lorenzo Leone, and Alioscia Hamma.
\newblock ``Magic-state resource theory for the ground state of the
  transverse-field ising model''.
\newblock \href{https://dx.doi.org/10.1103/PhysRevA.106.042426}{Phys. Rev. A
  {\bf 106}, 042426}~(2022).

\bibitem{odavić2022complexity}
Jovan Odavi{\'{c}}, Tobias Haug, Gianpaolo Torre, Alioscia Hamma, Fabio
  Franchini, and Salvatore~Marco Giampaolo.
\newblock ``Complexity of frustration: A new source of non-local
  non-stabilizerness''.
\newblock \href{https://dx.doi.org/10.21468/scipostphys.15.4.131}{{SciPost}
  Physics{\bf 15}}~(2023).

\bibitem{haug2023quantifying}
Tobias Haug and Lorenzo Piroli.
\newblock ``Quantifying nonstabilizerness of matrix product states''.
\newblock \href{https://dx.doi.org/10.1103/PhysRevB.107.035148}{Phys. Rev. B
  {\bf 107}, 035148}~(2023).

\bibitem{haug2023stabilizer}
Tobias Haug and Lorenzo Piroli.
\newblock ``Stabilizer entropies and nonstabilizerness monotones''.
\newblock \href{https://dx.doi.org/10.22331/q-2023-08-28-1092}{Quantum {\bf 7},
  1092}~(2023).

\bibitem{lami2023quantum}
Guglielmo Lami and Mario Collura.
\newblock ``Nonstabilizerness via perfect pauli sampling of matrix product
  states''.
\newblock \href{https://dx.doi.org/10.1103/PhysRevLett.131.180401}{Phys. Rev.
  Lett. {\bf 131}, 180401}~(2023).

\bibitem{tarabunga2023manybody}
Poetri~Sonya Tarabunga, Emanuele Tirrito, Titas Chanda, and Marcello Dalmonte.
\newblock ``Many-body magic via pauli-markov chains---from criticality to gauge
  theories''.
\newblock \href{https://dx.doi.org/10.1103/PRXQuantum.4.040317}{PRX Quantum
  {\bf 4}, 040317}~(2023).

\bibitem{tarabunga2023critical}
Poetri~Sonya Tarabunga.
\newblock ``Critical behaviours of non-stabilizerness in quantum spin
  chains''~(2023).
\newblock  \href{http://arxiv.org/abs/2309.00676}{arXiv:2309.00676}.

\bibitem{tarabunga2024nonstabilizerness}
Poetri~Sonya Tarabunga, Emanuele Tirrito, Mari~Carmen Bañuls, and Marcello
  Dalmonte.
\newblock ``Nonstabilizerness via matrix product states in the pauli
  basis''~(2024).
\newblock  \href{http://arxiv.org/abs/2401.16498}{arXiv:2401.16498}.

\bibitem{frau2024nonstabilizerness}
M.~Frau, P.~S. Tarabunga, M.~Collura, M.~Dalmonte, and E.~Tirrito.
\newblock ``Non-stabilizerness versus entanglement in matrix product
  states''~(2024).
\newblock  \href{http://arxiv.org/abs/2404.18768}{arXiv:2404.18768}.

\bibitem{chen2023magic}
Junjie Chen, Yuxuan Yan, and You Zhou.
\newblock ``Magic of quantum hypergraph states''~(2023).
\newblock  \href{http://arxiv.org/abs/2308.01886}{arXiv:2308.01886}.

\bibitem{oliviero2022measuring}
Salvatore F.~E. Oliviero, Lorenzo Leone, Alioscia Hamma, and Seth Lloyd.
\newblock ``Measuring magic on a quantum processor''.
\newblock \href{https://dx.doi.org/10.1038/s41534-022-00666-5}{npj Quantum
  Information {\bf 8}, 148}~(2022).

\bibitem{haug2023scalable}
Tobias Haug and M.S. Kim.
\newblock ``Scalable measures of magic resource for quantum computers''.
\newblock \href{https://dx.doi.org/10.1103/PRXQuantum.4.010301}{PRX Quantum
  {\bf 4}, 010301}~(2023).

\bibitem{haug2023efficient}
Tobias Haug, Soovin Lee, and M.~S. Kim.
\newblock ``Efficient stabilizer entropies for quantum computers''~(2023).
\newblock  \href{http://arxiv.org/abs/2305.19152}{arXiv:2305.19152}.

\bibitem{tirrito2023}
Emanuele Tirrito, Poetri~Sonya Tarabunga, Gugliemo Lami, Titas Chanda, Lorenzo
  Leone, Salvatore F.~E. Oliviero, Marcello Dalmonte, Mario Collura, and
  Alioscia Hamma.
\newblock ``Quantifying nonstabilizerness through entanglement spectrum
  flatness''.
\newblock \href{https://dx.doi.org/10.1103/physreva.109.l040401}{Physical
  Review A{\bf 109}}~(2024).

\bibitem{turkeshi2023measuring}
Xhek Turkeshi, Marco Schir\`o, and Piotr Sierant.
\newblock ``Measuring nonstabilizerness via multifractal flatness''.
\newblock \href{https://dx.doi.org/10.1103/PhysRevA.108.042408}{Phys. Rev. A
  {\bf 108}, 042408}~(2023).

\bibitem{amico2008}
Luigi Amico, Rosario Fazio, Andreas Osterloh, and Vlatko Vedral.
\newblock ``Entanglement in many-body systems''.
\newblock \href{https://dx.doi.org/10.1103/RevModPhys.80.517}{Rev. Mod. Phys.
  {\bf 80}, 517--576}~(2008).

\bibitem{eisert2010}
J.~Eisert, M.~Cramer, and M.~B. Plenio.
\newblock ``Colloquium: Area laws for the entanglement entropy''.
\newblock \href{https://dx.doi.org/10.1103/RevModPhys.82.277}{Rev. Mod. Phys.
  {\bf 82}, 277--306}~(2010).

\bibitem{bao2022}
Ning Bao, ChunJun Cao, and Vincent~Paul Su.
\newblock ``Magic state distillation from entangled states''.
\newblock \href{https://dx.doi.org/10.1103/PhysRevA.105.022602}{Phys. Rev. A
  {\bf 105}, 022602}~(2022).

\bibitem{Henley2004}
C~L Henley.
\newblock ``From classical to quantum dynamics at rokhsar{\textendash}kivelson
  points''.
\newblock \href{https://dx.doi.org/10.1088/0953-8984/16/11/045}{Journal of
  Physics: Condensed Matter {\bf 16}, S891--S898}~(2004).

\bibitem{Ardonne2004}
Eddy Ardonne, Paul Fendley, and Eduardo Fradkin.
\newblock ``Topological order and conformal quantum critical points''.
\newblock \href{https://dx.doi.org/10.1016/j.aop.2004.01.004}{Annals of Physics
  {\bf 310}, 493--551}~(2004).

\bibitem{Castelnovo2005}
Claudio Castelnovo, Claudio Chamon, Christopher Mudry, and Pierre Pujol.
\newblock ``From quantum mechanics to classical statistical physics:
  Generalized rokhsar{\textendash}kivelson hamiltonians and the
  {\textquotedblleft}stochastic matrix form{\textquotedblright}
  decomposition''.
\newblock \href{https://dx.doi.org/10.1016/j.aop.2005.01.006}{Annals of Physics
  {\bf 318}, 316--344}~(2005).

\bibitem{piemontese2023}
Stefano Piemontese, Tommaso Roscilde, and Alioscia Hamma.
\newblock ``Entanglement complexity of the rokhsar-kivelson-sign
  wavefunctions''.
\newblock \href{https://dx.doi.org/10.1103/PhysRevB.107.134202}{Phys. Rev. B
  {\bf 107}, 134202}~(2023).

\bibitem{Rokhsar_Kivelson}
Daniel~S. Rokhsar and Steven~A. Kivelson.
\newblock ``Superconductivity and the quantum hard-core dimer gas''.
\newblock \href{https://dx.doi.org/10.1103/PhysRevLett.61.2376}{Phys. Rev.
  Lett. {\bf 61}, 2376--2379}~(1988).

\bibitem{Verstraete2006}
F.~Verstraete, M.~M. Wolf, D.~Perez-Garcia, and J.~I. Cirac.
\newblock ``Criticality, the area law, and the computational power of projected
  entangled pair states''.
\newblock \href{https://dx.doi.org/10.1103/PhysRevLett.96.220601}{Phys. Rev.
  Lett. {\bf 96}, 220601}~(2006).

\bibitem{Schwarz2012}
Martin Schwarz, Kristan Temme, and Frank Verstraete.
\newblock ``Preparing projected entangled pair states on a quantum computer''.
\newblock \href{https://dx.doi.org/10.1103/PhysRevLett.108.110502}{Phys. Rev.
  Lett. {\bf 108}, 110502}~(2012).

\bibitem{zhu2019}
Guo-Yi Zhu and Guang-Ming Zhang.
\newblock ``Gapless coulomb state emerging from a self-dual topological
  tensor-network state''.
\newblock \href{https://dx.doi.org/10.1103/PhysRevLett.122.176401}{Phys. Rev.
  Lett. {\bf 122}, 176401}~(2019).

\bibitem{lee2022decoding}
Jong~Yeon Lee, Wenjie Ji, Zhen Bi, and Matthew P.~A. Fisher.
\newblock ``Decoding measurement-prepared quantum phases and transitions: from
  ising model to gauge theory, and beyond''~(2022).
\newblock  \href{http://arxiv.org/abs/2208.11699}{arXiv:2208.11699}.

\bibitem{zhu2022nishimoris}
Guo-Yi Zhu, Nathanan Tantivasadakarn, Ashvin Vishwanath, Simon Trebst, and
  Ruben Verresen.
\newblock ``Nishimori’s cat: Stable long-range entanglement from finite-depth
  unitaries and weak measurements''.
\newblock \href{https://dx.doi.org/10.1103/physrevlett.131.200201}{Physical
  Review Letters{\bf 131}}~(2023).

\bibitem{chen2023realizing}
Edward~H. Chen, Guo-Yi Zhu, Ruben Verresen, Alireza Seif, Elisa Baümer, David
  Layden, Nathanan Tantivasadakarn, Guanyu Zhu, Sarah Sheldon, Ashvin
  Vishwanath, Simon Trebst, and Abhinav Kandala.
\newblock ``Realizing the nishimori transition across the error threshold for
  constant-depth quantum circuits''~(2023).
\newblock  \href{http://arxiv.org/abs/2309.02863}{arXiv:2309.02863}.

\bibitem{zhu2023}
Guo-Yi Zhu, Ji-Yao Chen, Peng Ye, and Simon Trebst.
\newblock ``Topological fracton quantum phase transitions by tuning exact
  tensor network states''.
\newblock \href{https://dx.doi.org/10.1103/PhysRevLett.130.216704}{Phys. Rev.
  Lett. {\bf 130}, 216704}~(2023).

\bibitem{Castelnovo2010}
C.~Castelnovo, S.~Trebst, and M.~Troyer.
\newblock ``Fractionalization and topological order''.
\newblock In Understanding Quantum Phase Transitions.
\newblock \href{https://dx.doi.org/10.1201/b10273-10}{Pages 169--192}.
\newblock {CRC} Press~(2010).

\bibitem{Gross2021}
David Gross, Sepehr Nezami, and Michael Walter.
\newblock ``Schur{\textendash}weyl duality for the clifford group with
  applications: Property testing, a robust hudson theorem, and de finetti
  representations''.
\newblock \href{https://dx.doi.org/10.1007/s00220-021-04118-7}{Communications
  in Mathematical Physics {\bf 385}, 1325--1393}~(2021).

\bibitem{wolff1989}
Ulli Wolff.
\newblock ``Collective monte carlo updating for spin systems''.
\newblock \href{https://dx.doi.org/10.1103/PhysRevLett.62.361}{Phys. Rev. Lett.
  {\bf 62}, 361--364}~(1989).

\bibitem{Hukushima1996}
Koji Hukushima and Koji Nemoto.
\newblock ``Exchange monte carlo method and application to spin glass
  simulations''.
\newblock \href{https://dx.doi.org/10.1143/jpsj.65.1604}{Journal of the
  Physical Society of Japan {\bf 65}, 1604--1608}~(1996).

\bibitem{marinari1996}
E.~Marinari, G.~Parisi, J.~Ruiz-Lorenzo, and F.~Ritort.
\newblock ``Numerical evidence for spontaneously broken replica symmetry in 3d
  spin glasses''.
\newblock \href{https://dx.doi.org/10.1103/PhysRevLett.76.843}{Phys. Rev. Lett.
  {\bf 76}, 843--846}~(1996).

\bibitem{Wegner}
Franz~J. Wegner.
\newblock ``Duality in generalized ising models''~(2014).
\newblock  \href{http://arxiv.org/abs/1411.5815}{arXiv:1411.5815}.

\bibitem{Castelnovo2008}
Claudio Castelnovo and Claudio Chamon.
\newblock ``Quantum topological phase transition at the microscopic level''.
\newblock \href{https://dx.doi.org/10.1103/PhysRevB.77.054433}{Phys. Rev. B
  {\bf 77}, 054433}~(2008).

\bibitem{Talapov1996}
A~L Talapov and H~W~J Bl\"{o}te.
\newblock ``The magnetization of the 3d ising model''.
\newblock \href{https://dx.doi.org/10.1088/0305-4470/29/17/042}{Journal of
  Physics A: Mathematical and General {\bf 29}, 5727--5733}~(1996).

\bibitem{nishimori2001}
Hidetoshi Nishimori.
\newblock ``{Statistical Physics of Spin Glass and Information Processing: an
  Introduction}''.
\newblock
  \href{https://dx.doi.org/10.1093/acprof:oso/9780199227259.001.0001}{Oxford
  University Press}. ~(2001).

\bibitem{Sandvik2010}
Anders~W. Sandvik, Adolfo Avella, and Ferdinando Mancini.
\newblock ``Computational studies of quantum spin systems''.
\newblock In {AIP} Conference Proceedings.
\newblock {AIP}~(2010).

\bibitem{Wannier1950}
G.~H. Wannier.
\newblock ``Antiferromagnetism. the triangular ising net''.
\newblock \href{https://dx.doi.org/10.1103/PhysRev.79.357}{Phys. Rev. {\bf 79},
  357--364}~(1950).

\bibitem{Houtappel1950}
R.M.F. Houtappel.
\newblock ``Order-disorder in hexagonal lattices''.
\newblock \href{https://dx.doi.org/10.1016/0031-8914(50)90130-3}{Physica {\bf
  16}, 425--455}~(1950).

\bibitem{Blankschtein1984}
Daniel Blankschtein, M.~Ma, A.~Nihat Berker, Gary~S. Grest, and C.~M.
  Soukoulis.
\newblock ``Orderings of a stacked frustrated triangular system in three
  dimensions''.
\newblock \href{https://dx.doi.org/10.1103/PhysRevB.29.5250}{Phys. Rev. B {\bf
  29}, 5250--5252}~(1984).

\bibitem{Moessner2001}
R.~Moessner and S.~L. Sondhi.
\newblock ``Ising models of quantum frustration''.
\newblock \href{https://dx.doi.org/10.1103/PhysRevB.63.224401}{Phys. Rev. B
  {\bf 63}, 224401}~(2001).

\bibitem{Isakov2003}
S.~V. Isakov and R.~Moessner.
\newblock ``Interplay of quantum and thermal fluctuations in a frustrated
  magnet''.
\newblock \href{https://dx.doi.org/10.1103/PhysRevB.68.104409}{Phys. Rev. B
  {\bf 68}, 104409}~(2003).

\bibitem{Wang2017}
Yan-Cheng Wang, Yang Qi, Shu Chen, and Zi~Yang Meng.
\newblock ``Caution on emergent continuous symmetry: A monte carlo
  investigation of the transverse-field frustrated ising model on the
  triangular and honeycomb lattices''.
\newblock \href{https://dx.doi.org/10.1103/PhysRevB.96.115160}{Phys. Rev. B
  {\bf 96}, 115160}~(2017).

\bibitem{Katzgraber2006}
Helmut~G. Katzgraber, Mathias K\"orner, and A.~P. Young.
\newblock ``Universality in three-dimensional ising spin glasses: A monte carlo
  study''.
\newblock \href{https://dx.doi.org/10.1103/PhysRevB.73.224432}{Phys. Rev. B
  {\bf 73}, 224432}~(2006).

\bibitem{CJMM22}
Jonas Charfreitag, Michael J{\"u}nger, Sven Mallach, and Petra Mutzel.
\newblock ``{M}c{S}parse: {E}xact solutions of sparse maximum cut and sparse
  unconstrained binary quadratic optimization problems''.
\newblock In Cynthia~A. Phillips and Bettina Speckmann, editors, 2022
  Proceedings of the Symposium on Algorithm Engineering and Experiments
  ({ALENEX}).
\newblock \href{https://dx.doi.org/10.1137/1.9781611977042.5}{Pages 54--66}.
\newblock ~(2022).

\bibitem{Kitaev2006}
Alexei Kitaev.
\newblock ``Anyons in an exactly solved model and beyond''.
\newblock \href{https://dx.doi.org/10.1016/j.aop.2005.10.005}{Annals of Physics
  {\bf 321}, 2--111}~(2006).

\bibitem{Castelnovo2007}
Claudio Castelnovo, Claudio Chamon, Christopher Mudry, and Pierre Pujol.
\newblock ``Zero-temperature kosterlitz{\textendash}thouless transition in a
  two-dimensional quantum system''.
\newblock \href{https://dx.doi.org/10.1016/j.aop.2006.04.017}{Annals of Physics
  {\bf 322}, 903--934}~(2007).

\bibitem{Moessner2003}
R.~Moessner and S.~L. Sondhi.
\newblock ``Three-dimensional resonating-valence-bond liquids and their
  excitations''.
\newblock \href{https://dx.doi.org/10.1103/PhysRevB.68.184512}{Phys. Rev. B
  {\bf 68}, 184512}~(2003).

\bibitem{Hermele2004}
Michael Hermele, Matthew P.~A. Fisher, and Leon Balents.
\newblock ``Pyrochlore photons: The $u(1)$ spin liquid in a $s=\frac{1}{2}$
  three-dimensional frustrated magnet''.
\newblock \href{https://dx.doi.org/10.1103/PhysRevB.69.064404}{Phys. Rev. B
  {\bf 69}, 064404}~(2004).

\bibitem{Castro2006}
A.~H. Castro~Neto, P.~Pujol, and Eduardo Fradkin.
\newblock ``Ice: A strongly correlated proton system''.
\newblock \href{https://dx.doi.org/10.1103/PhysRevB.74.024302}{Phys. Rev. B
  {\bf 74}, 024302}~(2006).

\bibitem{Benton2012}
Owen Benton, Olga Sikora, and Nic Shannon.
\newblock ``Seeing the light: Experimental signatures of emergent
  electromagnetism in a quantum spin ice''.
\newblock \href{https://dx.doi.org/10.1103/PhysRevB.86.075154}{Phys. Rev. B
  {\bf 86}, 075154}~(2012).

\end{thebibliography}

\end{document}